\documentclass[pre,amsmath]{revtex4}
\usepackage{graphicx}
\usepackage{color}
\usepackage{bm}
\usepackage{epsfig}
\usepackage{latexsym}
\usepackage{pifont}
\usepackage{float}
\graphicspath{{figure/}}
\usepackage[utf8]{inputenc}
\usepackage{siunitx}

\begin{document}

\today

\title{Modelling cellular spreading and emergence of motility in the presence of curved membrane proteins and active cytoskeleton forces}

\author{Raj Kumar Sadhu$^1$, Samo Peni\v{c}$^2$, Ale\v{s} Igli\v{c} $^{2,3}$ and Nir S. Gov$^1$}
\affiliation{$^1$ Department of Chemical and Biological Physics, Weizmann Institute, Rehovot, Israel}
\affiliation{$^2$ Faculty of Electrical Engineering, University of Ljubljana, Ljubljana, Slovenia}
\affiliation{$^3$ Faculty of Medicine, University of Ljubljana, Ljubljana, Slovenia}

\begin{abstract}
Eukaryotic cells adhere to extracellular matrix during the normal development of the organism, forming static adhesion as well as during cell motility. We study this process by considering a simplified coarse-grained model of a vesicle that has uniform adhesion energy with a flat substrate, mobile curved membrane proteins and active forces. We find that a high concentration of curved proteins alone increases the spreading of the vesicle, by the self-organization of the curved proteins at the high curvature vesicle-substrate contact line, thereby reducing the bending energy penalty at the vesicle rim. This is most significant in the regime of low bare vesicle-substrate adhesion. When these curved proteins induce protrusive forces, representing the actin cytoskeleton, we find efficient spreading, in the form of sheet-like lamellipodia. Finally, the same mechanism of spreading is found to include a minimal set of ingredients needed to give rise to motile phenotypes.
\end{abstract}
\maketitle

\section{Introduction}
\label{intro}
The adhesion of cells to an external substrate is an essential process allowing cells to form cohesive tissues, migrate and proliferate \cite{geiger2009environmental}. The stages of cellular spreading over an adhesive surface have been studied experimentally \cite{Sheetz2004PRL,cavalcanti2007cell,cuvelier2007universal,Sheetz2008,Sheetz2011,Sheetz2014,schaufler2016selective,Nils2017}, and involve an initial stage of non-specific and weak adhesion, followed usually by spreading that is driven by the formation of thin sheet-like lamellipodia. These structures form when actin polymerization is recruited to the leading edge of the lamellipodia \cite{Oakes2018}. The actin provides both a protrusive force that pushes the membrane outwards and traction forces that enhance the growth of adhesion complexes. When actin polymerization is inhibited, cells exhibit very weak spreading, small adhered area \cite{Sheetz2008}, and strongly retract (if the drug is delivered after normal spreading \cite{bar1999pearling}).

While this complex process has been explored from its biological aspects, a more basic physics understanding of the cell spreading process is lacking \cite{schwarz2013physics}. There are several theoretical treatments of the cell spreading process, with various levels of coarse-graining and detail, starting from the simplest dynamical-scaling model \cite{cuvelier2007universal}. Some models focus on the role of actin and actin-adhesion coupling during the spreading and adhesion, but do not describe the membrane shape dynamics in detail \cite{Sheetz2010,nisenholz2014active,gong2018matching,dedenon2019model}. Other models describe in detail the cell shape and the stress-fibers that span the adhered cell \cite{loosli2010cytoskeleton,fang2020active}, with the higher realism obtained at a price of much higher model complexity.

The simpler process of vesicle adhesion, which has been explored using in-vitro systematic experiments \cite{gruhn2007novel,reister2008dynamics,streicher2009integrin,maan2018adhesion,bibissidis2020}, is amenable to theoretical physics description \cite{bibissidis2020}. A large number of theoretical studies treated the coarse-grained adhesion of a vesicle with uniform adhesion \cite{lipowski1990,lipowsky1991,lipowski2005}, while other studies have explored the molecular-scale adhesion dynamics \cite{boulbitch2001kinetics,smith2008force,sengupta2018adhesion}, which include ligand binding/unbinding as well as diffusion and aggregation of ligands on the membrane-substrate interface (through direct and membrane-induced interactions \cite{farago2010fluctuation,fenz2017membrane}).

We aim here to help bridge the gap between our understanding of vesicle adhesion and the more complex process of active cellular spreading. We do this by exploring a simple, coarse-grained theoretical model of vesicle adhesion which contains two ingredients: (i) we add a fixed density of curved membrane proteins, and (ii) exert active protrusive forces at the locations of the membrane proteins. Both of these components are motivated by experimental properties of cells: cell membranes contain a plethora of curved membrane proteins \cite{zimmerberg2006proteins,suetsugu2014dynamic}, and many of these curved proteins (or membrane-bound protein complexes) are involved in the recruitment of actin polymerization activity to the membrane \cite{scita2008irsp53,kuhn2015structure}. These two ingredients have been recently shown theoretically \cite{miha2019} and experimentally \cite{begemann2019mechanochemical,graziano2019cell} to be sufficient to induce the formation of sheet-like lamellipodia protrusions in cells. We therefore set out to explore theoretically the role of these ingredients during active spreading of cells.

\section{Model}
\label{sec:model}
\begin{figure}[ht]
\centering
\includegraphics[scale=0.7]{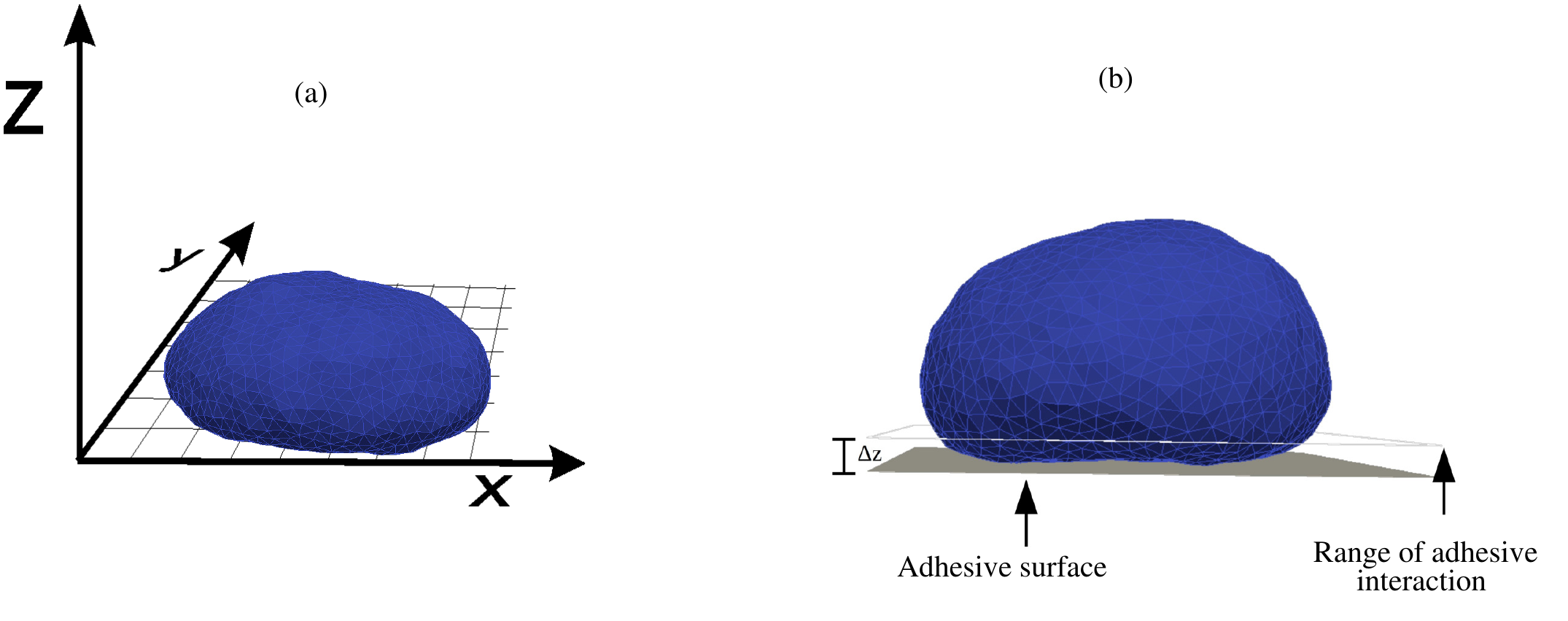}
\caption{A schematic representation of our model. (a) A three-dimensional vesicle is placed on the adhesive surface, having uniform adhesion interaction with the vesicle throughout. The position of the adhesive surface is at $z=z_{ad}$ and parallel to $x-y$ plan. (b) The range of adhesive interaction is within a distance of $\Delta z$ above the adhesive surface.  The total adhesion energy will be $E_{ad}$ times the number of vertices within this interaction range, where $E_{ad}$ is the adhesion strength, defined as the adhesion energy per adhered vertex.}
\label{model} 
\end{figure}

We use here the same theoretical model of \cite{miha2019}, adapted to include membrane-substrate adhesion. We consider a three-dimensional vesicle that is described by a surface of $N$ vertices, each connected to its neighbours with bonds of length $l$, to form closed, dynamically triangulated, self-avoiding network, with the topology of a sphere, as shown in Fig. \ref{model}. An adhesive surface is placed near the vesicle, parallel to x-y plane and at position $z=z_{ad}$ (see Fig. \ref{model}). The total energy of the vesicle is the sum of four contributions, (1) the local bending energy due to its curvature, (2) the energy due to binding between neighboring proteins (direct interaction energy), (3) the energy due to the active cytoskeleton force and (4) the adhesive energy due to the attractive interaction between the vesicle and the substrate. 

Note that the term "curved membrane proteins" stands for any complex of such proteins and lipids (such as in nanodomains) in general, that has a spontaneous curvature, and can induce local polymerization of the cortical actin cytoskeleton.

The bending energy can be mathematically expressed using the Helfrich expression  \cite{helfrich1973} as,

$$W_b=\frac{\kappa}{2} \int_A (C_1 + C_2 - C_0)^2 dA $$
where, $C_1$ and $C_2$ are principle curvatures, $C_0$ is the spontaneous curvature at any position of the vesicle, and $\kappa$ is the bending rigidity. The bending energy is properly discretized  following the refs. \cite{gompper1996,ramakrishnan2011,samo2015}. We model the spontaneous curvature as discrete entities that is occupied by a vertex. The spontaneous curvature of a vertex that is occupied by curved proteins is taken to have some non-zero value $C_0=c_0$, and zero otherwise.  In our model, we consider a positive $c_0>0$, i.e. convex spontaneous curvature. Note that we describe here isotropic curved proteins or isotropic curved nanodomains \cite{kralj_iglic1996,fovsnarivc2006influence}.

The energy due to the binding between proteins is expressed as,

$$W_d = -w\sum_{i<j} {\cal H} (r_0 - r_{ij})$$

where, ${\cal H}$ is the Heaviside step function, having a value of unity if the argument is positive, otherwise vanishes, $r_{ij}$ is the displacement between proteins and $r_0$ is the range of attraction, beyond which the attractive force becomes zero and $w$ is a positive constant (throughout the paper we use: $w=1 ~k_BT$). In our model, we choose $r_0$ to be such that only the proteins in neighbouring vertices can bind with each other.

The actin cytoskeleton that is recruited by the curved proteins exerts an outward force, which therefore gives the following energy contribution,

$$W_F = -F \sum_i \hat{n_i}.\overrightarrow{x_i}$$ 

where, $F$ is the magnitude of the force, $\hat {n_i}$ be the outward normal of the vertex that contains a protein and $\overrightarrow{x_i}$ is the position vector of the protein. Due to the thermal fluctuations of the vesicle and the diffusion of the proteins, this force changes its direction with time. Thus, this is equivalent to a force exerted due to a time-varying external potential, which therefore drives the system out of equilibrium.

Finally, the vesicle can adhere on the adhesive surface, due to which it has the energy contribution,

$$W_A = -\int_{A} V(z) dA $$

where, $V(z)$ is the interaction potential between the adhesive surface and  the vesicle. In our model, we choose the interaction potential $V(z)$ to be a step potential, such that, $V(z) = E_{ad}$ (a constant, termed as adhesion strength) for  $z_{ad} \leq z(i) \leq (z_{ad} + \Delta z)$, and zero otherwise; where $z_{ad}$ is the z-coordinate of the adhesive surface, $\Delta z$ is the width of potential energy and $z(i)$ is the z-coordinate of $i$-th vertex (see Fig. \ref{model} (b)). The adhesive surface is considered to be a rigid object, such that a vertex can not penetrate it.

Thus, the total energy of the system can be written as,

\begin{equation}
W = W_b + W_d + W_F + W_A
\label{eq:energy}
\end{equation}

The simulation details are given in the appendix \ref{sec:simulation_detail}. Throughout these simulations we do not conserve the vesicle volume, which is appropriate for cells that are observed to change their volume significantly during spreading and adhesion \cite{guo2017cell,xie2018controlling}. One can however add to the model the effects of an internal osmotic pressure that inflates the vesicle \cite{miha2019}.

Note that during the process of vesicle adhesion and spreading, there are hydrodynamic processes that we do not include in our model, such as fluid flow within the vesicle, between the vesicle and the substrate, and of the fluid membrane, i.e. visco-elastic properties of the system were neglected  (see for example \cite{bernard2000strong}). These omissions mean that the dynamics that we extract from the simulations, in MC time steps, may not be simply mapped to a real time-scale.
 
\section{Results} 
In order to validate our simulation method, we first compared the steady-state shapes of adhered protein-free vesicles to those previously obtained using detailed numerical solutions \cite{raval2020shape,bibissidis2020}. The very good agreement between the two different methods (Fig. \ref{compare_jeel} of appendix \ref{sec:jeel}) verifies the accuracy of our simulations. The equilibrium vesicle shapes minimize the energy, striking a balance between the adhesion energy that drives the spreading and the bending energy that resists the deformation of the membrane.

Next, we explored the effects of the curved proteins and the active forces that they recruit.

\label{sec:results}
\subsection{Spreading of vesicles with passive curved proteins}

We start by exploring the effects of passive curved proteins, without active cytoskeletal forces.  Since, there is no active force acting on the system, the system reaches an  equilibrium configuration after evolving it for sufficiently long times. We aim to understand here, how the presence of the curved proteins affects the vesicle shape, as well as the demixing of the curved proteins. 

\begin{figure}[ht]
\centering
\includegraphics[scale=2.1]{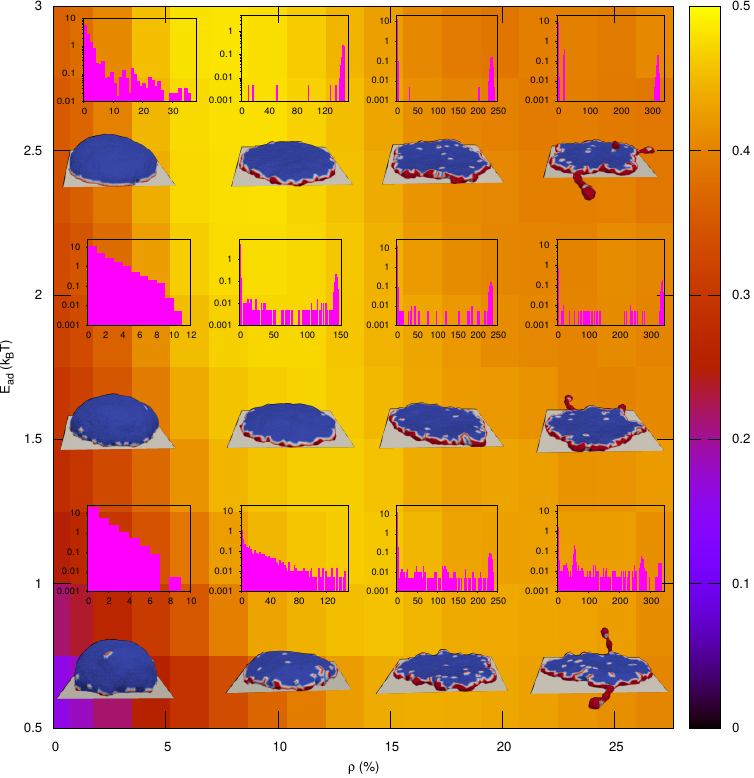}
\caption{Typical steady state configurations of the vesicles and cluster distribution in the $E_{ad}-\rho$ plane for the passive case (F=0). The background color is showing the fraction of adhered area of the vesicle on the adhesive surface. The blue part in the vesicle  denotes the protein free regions and the red color denotes the curved proteins. In the inset, we show the cluster distribution of proteins near each snapshot; x-axis is the size of the cluster and the y-axis is  the corresponding frequency of having the particular cluster in the ensemble. The y-axis is shown in the log-scale.  We show the snapshots for $E_{ad}$=0.75, 1.5, 2.5 (in units of $k_B T$ ) and $\rho$=3.45 \%, 10.36 \%, 17.27 \% and 24.18 \%. Other parameters are: Total number of vertices, $N=1447$, $\kappa =20~ k_B T$ and $w=1 ~k_BT$. The width of the potential, $\Delta z$ is taken to be $l_{min}$, and the spontaneous curvature of the curved proteins is taken to be $c_0=1~l^{-1}_{min}$.}
\label{phase_passive} 
\end{figure}

In Fig. \ref{phase_passive}, we plot snapshots of typical equilibrium adhered vesicle shape and the protein cluster-size distribution for different values of adhesion strength ($E_{ad}$) and the number density of proteins ($\rho=N_c/N$). The background color is showing the adhered area fraction (AAF, $A_{ad}/A$), where $A_{ad}$ is the area adhered on the adhesive surface and $A$ is the total area of the vesicle (so that the maximal possible value is $A_{ad}/A\simeq0.5$ for very small volume). 

\begin{figure}[ht]
\centering
\includegraphics[scale=1]{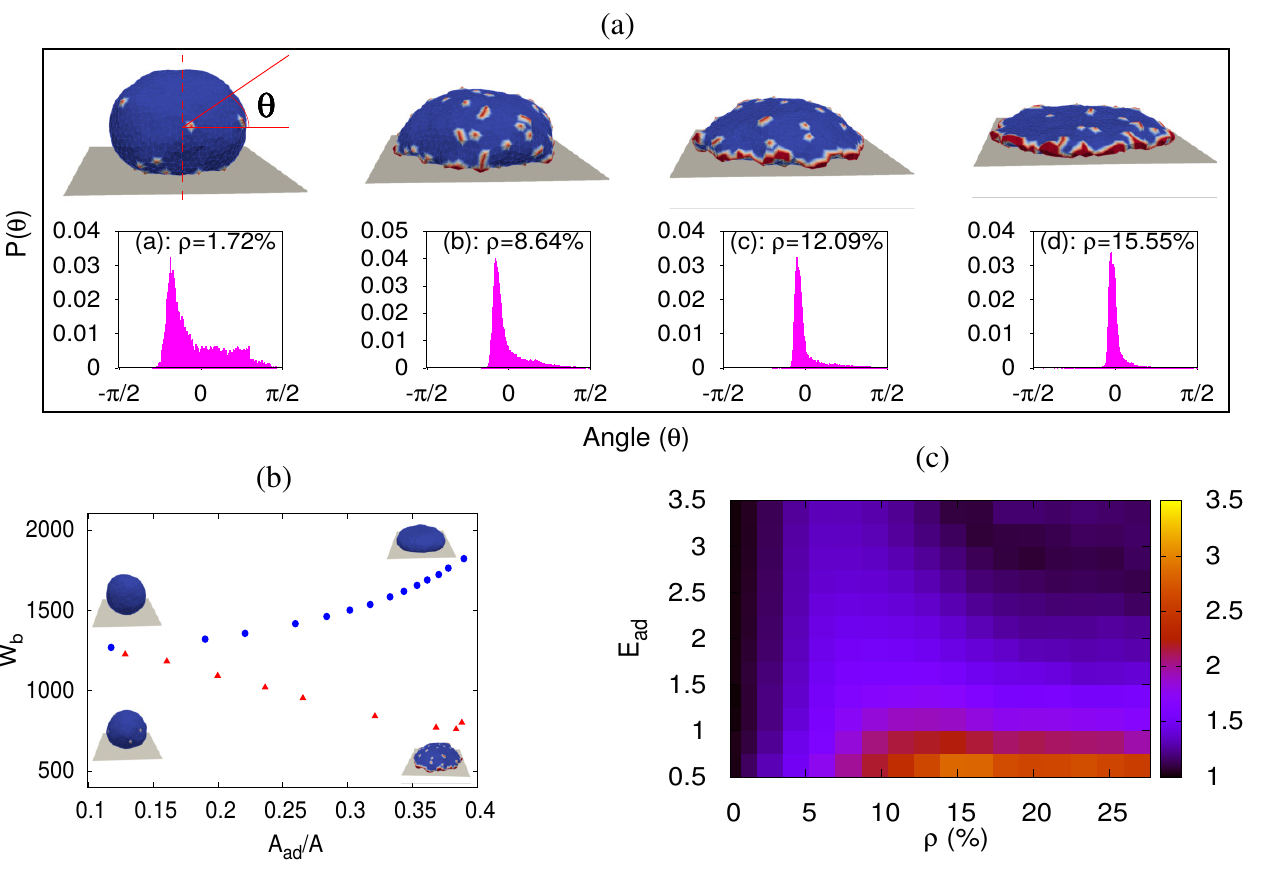}
\caption{Curved proteins facilitate spreading at low adhesion energy. (a) Protein distribution along the angle $\theta$ for the passive case with $E_{ad}=0.50 ~k_B T$ and various $\rho$ along with the snapshots. The angle ($\theta$) is defined as the angle between the line parallel to the x-y plane, passing through the COM of vesicle and the line joining the surface of the vesicle and its COM, as shown in figure. P($\theta$) is the probability that there is a protein at the angle $\theta$, averaged over the azimuthal direction, such that $\theta$ varies from $-\pi/2$ to $\pi/2$.  (b) Bending energy as a function of AAF with and without proteins. The blue circles are for protein-free vesicle and the red triangles are for passive case. We also show the snapshots for each case, in the minimum and the maximum adhered state. For protein-free vesicle, we vary $E_{ad}$ from $0.50 ~k_B T$ to $4.0~ k_B T$, while for passive vesicle, we fix $E_{ad}$ to $0.50 ~k_B T$ and vary $\rho$ from $0.70 ~\%$ to $27.64 ~\%$.  (c) Ratio of AAF between vesicles with highly curved passive proteins (Fig.\ref{phase_passive}) and protein-free vesicles. The ratio is close approaches unity for either $\rho\rightarrow 0$ or for $E_{ad}\rightarrow\infty$.}
\label{passive_cluster} 
\end{figure}

For small $\rho$ and small $E_{ad}$, the vesicle shape is quasi-spherical, similar to adhered protein-free vesicles. The proteins are weakly clustered and there are small patches of proteins around the whole vesicle. As $\rho$ increases keeping $E_{ad}$ small, the vesicle spreads, becomes flatter (since the volume is not fixed). This spreading is driven by the aggregation of the curved proteins at the high-curvature region \cite{kralj_iglic1996,markin1981} where the adhered vesicle contacts the substrate ("contact line", Fig.\ref{passive_cluster} (a)). Due to their large convex spontaneous curvature, the aggregation of the proteins in this region lowers significantly the bending energy of the vesicle, facilitates stronger bending and larger spreading on the substrate compared to the protein-free vesicle (Fig.\ref{passive_cluster} (b)). 

Note that demixing and phase-separation of membrane proteins on an adhered vesicle were considered theoretically \cite{rouhiparkouhi2013adhesion}, however the curvature-based demixing \cite{aimon2014membrane} that we discuss here was not previously treated during membrane adhesion. A simplified analytic model (see appendix \ref{sec:analytical}) allows us to qualitatively recover the same trends found in the simulations. We find that the aggregation of curved proteins along the high curvature rim of the vesicle leads to a monotonous increase in the AAF and decrease of the bending energy with increasing protein density (for low $\rho$, Fig. \ref{passive_analytical}).

As $\rho$ increases, the vesicle flattens and surplus proteins that have no more space along the contact line form necklace-like protein clusters around the vesicle (Fig. \ref{phase_passive}). The necklace-like structures are formed because of the isotropic membrane proteins used in our simulation \cite{miha2019,Luka2017}. 
The AAF increases with $\rho$ as long as the shape of the vesicle remains quasi-spherical or pancake-like. However, for large $\rho$, the shape of the vesicle changes and the surplus curved proteins lead to budding all over the membrane. These buds (isolated and necklace-like), deform the membrane away from the flat shape, and give rise to a decrease of the AAF with increasing $\rho$ (Fig. \ref{phase_passive}). 

As $E_{ad}$ is increased, the vesicle is more spread, and the natural curvature along the contact line increases. This has the effect of aggregating the curved proteins more strongly and the flattening effect of the curved proteins sets in at lower densities (Fig. \ref{phase_passive}). Similarly to low $E_{ad}$, the AAF decreases for large $\rho$, and the peak in the AAF shifts to smaller values of $\rho$. As $E_{ad}$ increases the proteins form a large cluster at the contact line, where almost all the proteins are clustered in a single cluster.

To quantify the effect of the curved proteins on the vesicle spreading, we measure the ratio of AAF between vesicles with highly curved passive proteins and protein-free vesicles (Fig.\ref{passive_cluster} (c)). We find that this ratio is larger for small $E_{ad}$, where the protein-free vesicles are very weakly adhered. As expected from Fig. \ref{phase_passive}, the enhancement of spreading due to the curved proteins has a maximum as function of $\rho$, with the peak shifted to lower values of $\rho$ as $E_{ad}$ increases. Note that for the passive case ($F=0$), a protein-free vesicle is similar to a vesicle with flat proteins ($c_0=0$). For details, please see appendix \ref{sec:passive_c0_0}.

\subsection{Spreading of vesicles with active curved proteins}
\begin{figure}[ht]
\centering
\includegraphics[scale=2.1]{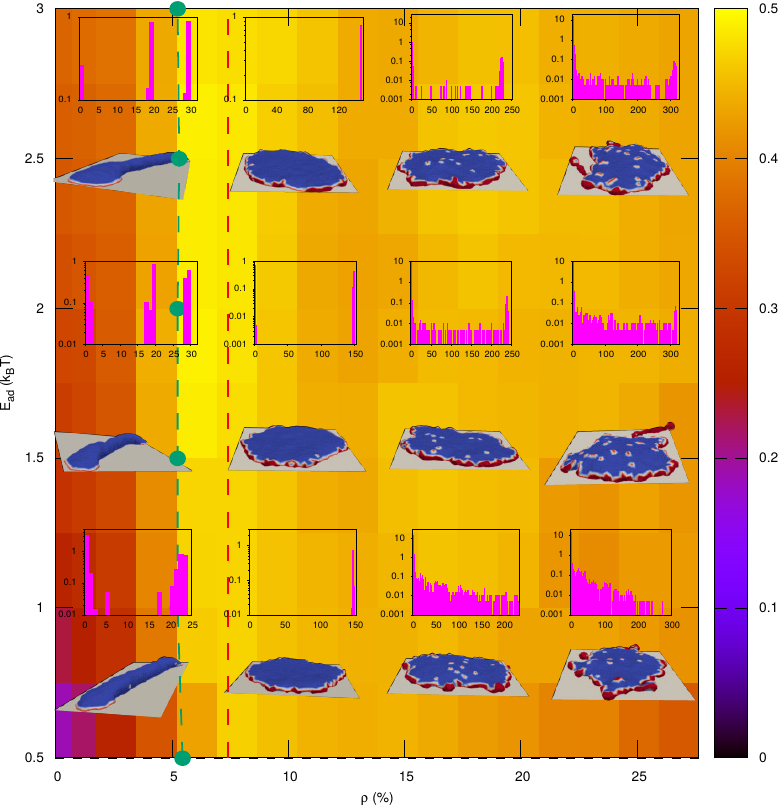}
\caption{Typical steady-state configuration of vesicle shape and cluster-size distribution with active forces. Similar to Fig. \ref{phase_passive} for an active protrusive force $F=4 ~k_B T/l_{min}$. The green dashed line-circles denotes the transition to a pancake-like shape. The red dashed vertical line denotes the density for the pancake transition for a free vesicle (without adhesion, \cite{miha2019}). We estimate these transition lines by measuring the mean cluster size, which shows a sharp jump at the critical $\rho$ (Fig. \ref{phase_active_pancake}(c)). Snapshots are shown for $E_{ad}$=0.75, 1.5, 2.5 (in units of $k_B T$) and $\rho=3.45 ~\%, 10.36 ~\%, 17.27 ~\%$ and $24.18 ~\%$. All the other parameters are the same as in Fig. \ref{phase_passive}. }
\label{phase_active} 
\end{figure}

Next, we study the active system, with active cytoskeletal forces acting outward on the proteins. To highlight the effects of the active force, we start with a large value of $F= 4 ~k_BT/l_{min}$ (Fig. \ref{phase_active}). Despite the presence of the active force, we find that for sufficiently large times most systems do reach a well-defined steady state, which allows us to extract average quantities, such as the AAF and the cluster distribution. When the density $\rho$ is small, the shapes of the adhered vesicle are quite different compared to the passive case (Fig. \ref{phase_passive}). However, for large $\rho$, the shapes are quite similar to the passive case. 

For small value of $\rho$, unlike the passive case (Fig. \ref{phase_passive}), the shape of the vesicle is highly non-rotationally-symmetric (Fig. \ref{phase_active}). The transition into this class of shapes, for $\rho$ to the left of the vertical dashed green line (around $5.25 ~\%$), is very sharp and was also observed for free vesicles with active curved proteins (denoted by the vertical dashed red line, $\rho \sim 7.26 ~\%$) \cite{miha2019}. At a low number of proteins, there are simply not enough proteins to complete a circular aggregate around the rim of the adhered vesicle. Instead, the proteins settle into two opposing arc-like aggregates, which exert opposing forces on the vesicle, that therefore assumes a stretched tube-like shape. 

\begin{figure}[ht]
\centering
\includegraphics[scale=0.7]{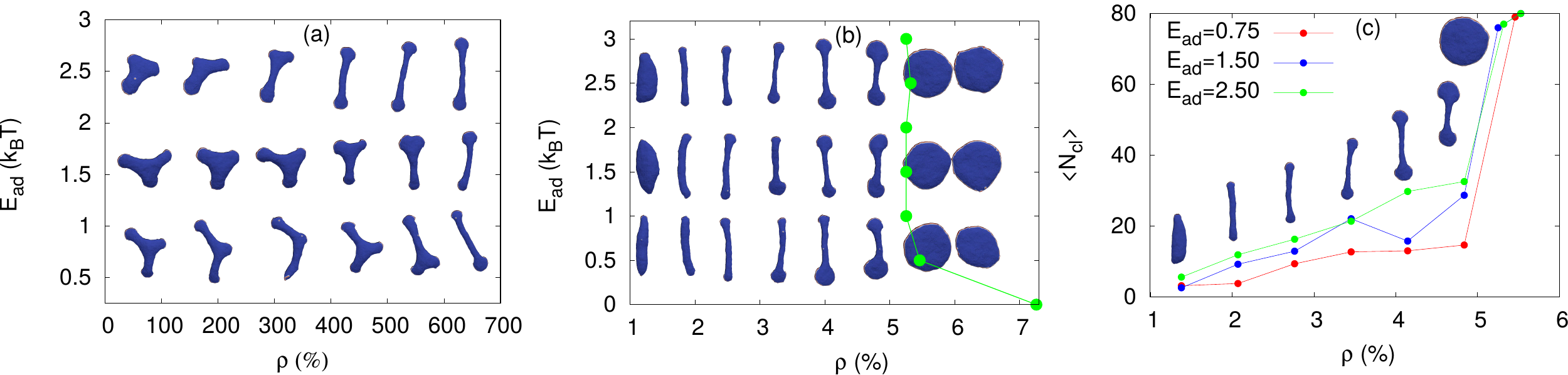}
\caption{Transition to pancake-like, highly-spread shape, for active curved proteins ($F=4 ~k_B T/l_{min}$).  (a) Snapshots of the vesicle for small density, at different instant of time. Here, we show that there are different possible metastable states in the small density regime. We use here $\rho=3.45 ~\%$, $E_{ad}= 0.75, 1.5, 2.5$ (in units of $k_B T$). (b) Configurations of vesicle for different values of $E_{ad}$ and $\rho$. The green line with circles denotes the density for a given $E_{ad}$ at which the pancake-like shape is obtained. (c) Mean cluster size of proteins, $\langle N_{cl}\rangle$, as a function of $\rho$ for different values of $E_{ad}$. The sharp increase in the value of $\langle N_{cl}\rangle$ shows the discontinuous transition to pancake-like shape. We also show the snapshots of vesicle for $E_{ad}=2.5~ k_B T$ for different densities to show how the jump in the value of $\langle N_{cl}\rangle$ gives rise to the pancake-like shape transition.}
\label{phase_active_pancake} 
\end{figure}

In Fig. \ref{phase_active_pancake}(a) we demonstrate the dynamics of the spreading in the low $\rho$ regime. We find that often the proteins form three or more aggregates that drive the spreading, but they coarsen over time to form the stable two-arc shapes. These two-arc shapes are mostly non-motile, but due to asymmetry between the sizes of the arc-like protein aggregates on either side, there can be a net force that leads to sliding of these vesicles in the direction of the end that has the larger aggregate (see Fig.\ref{phase_crescent}(c)). 

For $\rho$ values to the right of the red dashed line in Fig. \ref{phase_active} there is a transition to pancake-like shapes, that correspond to very efficient spreading and high AAF. For even larger protein density, there is not enough space on the outer rim of the vesicle to accommodate all the proteins, so the pancake-like shape does not remain stable. Small bud like structure appear around the vesicle and the AAF decreases, similar to the passive case at high densities (Fig. \ref{phase_passive}).

The transition between the two-arc and pancake-like shapes for different values of $\rho$ and $E_{ad}$ is shown in more detail in Fig. \ref{phase_active_pancake}(b). As $E_{ad}\rightarrow0$, the transition density increases and approaches its value for a free vesicle. The pancake transition is quantified by the mean cluster size ($\langle N_{cl}\rangle$), which we plot as function of $\rho$ (Fig. \ref{phase_active_pancake}(c)). The quantity $\langle N_{cl}\rangle$ exhibits a sharp jump near the pancake transition, where one cluster contains almost all the proteins (phase separation).

\begin{figure}[ht]
\centering
\includegraphics[scale=0.7]{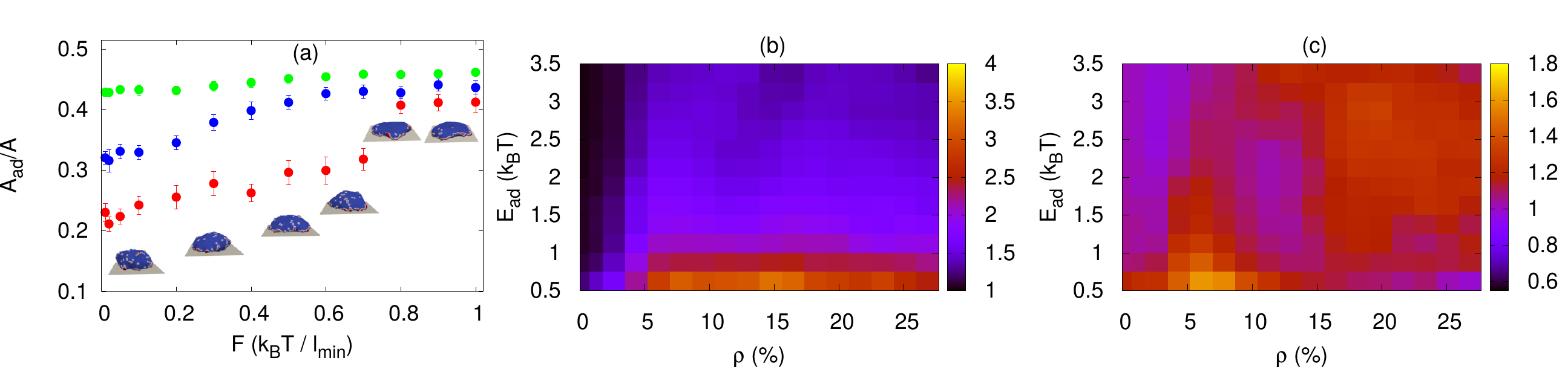}
\caption{Variation of adhered fraction due to the active force ($F$). (a) Fraction of adhered area with $F$ for different values of $E_{ad}$ for a fixed $\rho$ ($=10.36~\%$). For small $E_{ad}$, the fraction $A_{ad}/A$ increases with $F$ significantly, however, as $E_{ad}$ increases, the value of  $A_{ad}/A$ does not vary much. We also show the shape of the vesicle for $E_{ad}=0.25 ~k_B T$. (b)  The ratio of adhered area of active vesicle with $c_0=1~l^{-1}_{min}$ to the passive vesicle with $c_0=0$. The ratio is maximal for small $E_{ad}$ over a wide range of $\rho$. (c) The ratio of adhered area of active to passive vesicle, with spontaneous curvature $c_0=1~l^{-1}_{min}$. We note that the maximum increase in the adhered area of the active vesicle over passive one is in the small $\rho$ and small $E_{ad}$ region.}
\label{ad_area_active} 
\end{figure}

\begin{figure}[ht]
\centering
\includegraphics[scale=1.9]{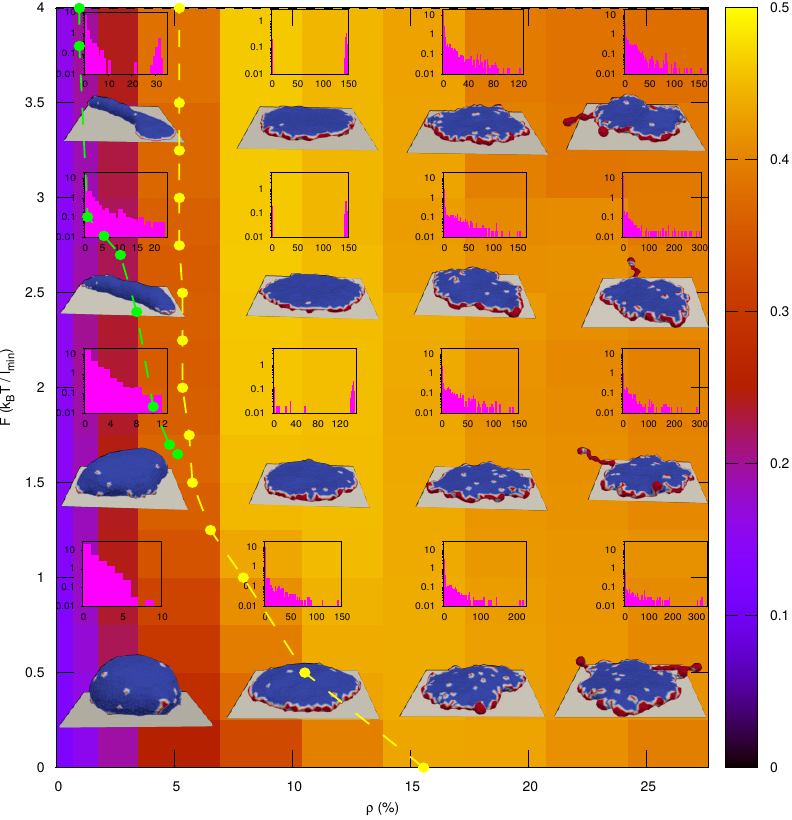}
\caption{Typical steady-state configurations of vesicle shape and cluster-size distribution for different $F$  and $\rho$, for a given small adhesion strength: $E_{ad}=0.50 ~k_B T$. The snapshots are shown for $F$=0.50, 1.50, 2.50 and 3.50 (in units of $k_B T/l_{min}$), and $\rho=3.45 ~\%, 10.36 ~\%, 17.27 ~\%$ and $24.18 ~\%$. The  yellow dotted lines denotes the transition to a pancake-like shape. For large $F$, the transition to a pancake shape from a two-arc shape is very sharp, and the transition line is estimated by measuring the mean cluster, as explained in Fig. \ref{phase_active_pancake}. For small $F$, the transition is not very sharp. Here, in order to identify the transition line, we measure the largest cluster, and we approximate the shape to be a pancake when the largest cluster is at least $60 ~\%$ of the total number of proteins (see Fig. \ref{largest_cluster} of appendix \ref{sec:largest_cluster}). The green dotted line represents the transition from a quasi-spherical to a two-arc type shape. The slope of this transition line diverges as $\rho \rightarrow 0$. This transition is also estimated by measuring the largest cluster and the threshold value of the largest cluster is taken to be $30 ~\%$ in this case, as there are two separate clusters in the two-arc shape. All the other parameters are the same as in Fig. \ref{phase_passive}. }
\label{phase_F_rho} 
\end{figure}

Next, we study the AAF as a function of $F$, for different values of $E_{ad}$ and a fixed density $\rho=10.36~\%$ (Fig. \ref{ad_area_active} (a)). For large $E_{ad}$ the force increases the AAF smoothly, as the vesicles are already spread even in the absence of active forces. For small $E_{ad}$, the AAF shows a large increase with $F$, including an abrupt jump for $E_{ad}=0.25 ~k_B T$ at $F \sim 0.8 ~k_B T/l_{min}$. This jump corresponds to crossing the pancake transition line for these parameters. In Fig. \ref{phase_F_rho} we plot a more complete phase-diagram, of the steady-state shapes of the vesicle for low adhesion ($E_{ad}=0.5 ~k_B T$), as function of the proteins density and active force. The active force is seen to shift the transition to the pancake (or the two-arc) shape, to lower values of the density, compared with the passive system (at $F=0$). This shift to lower densities due to the force is also seen for the initiation of protein aggregation (appendix \ref{sec:Ncl2line}, Fig.\ref{Fc_linear_stability}).

The role of the active force in increasing the steady-state AAF is emphasized in Fig. \ref{ad_area_active} (b), where we plot its ratio with the AAF of the protein-free vesicle. We find that the largest increase in AAF due to the active curved proteins is for low $E_{ad}$. Compared to the vesicle containing curved passive proteins, this enhanced spreading is extended to lower values of $\rho$ due to the pancake transition (compare to Fig. \ref{passive_cluster}(c)). This is emphasized in Fig. \ref{ad_area_active} (c) where we plot the ratio of the AAF of the active and passive curved protein systems. The largest contribution of the active force is for low $E_{ad}$ and $\rho$, where the passive proteins do not form strong aggregation at the contact line and are ineffective in driving spreading, while the added active forces drive the pancake transition and strongly enhanced spreading.

We note that for large $E_{ad}$ and large $\rho$, the AAF is also increased due to activity, compared to the passive curved proteins (Fig. \ref{ad_area_active} (c)). This is the region where the passive proteins form large necklace-like structures which decrease the adhered area (Fig. \ref{phase_passive}), while the active forces tend to destabilize them and therefore increase the AAF. 

The active forces exerted at the locations of the curved proteins may give rise to a non-zero net force. While the planar component of this force simply pushes the vesicle on the substrate, the vertical component (along $z$-direction) can affect the adhesion. Since the curved proteins prefer the free (dorsal) side of the vesicle over the perfectly flat basal side, this force tends to overall push the vesicle away from the substrate.
In the regime of very low $E_{ad}$ and large $\rho$, the active forces exerted by the proteins can lead to lowering the AAF by partially detaching the vesicle. In Fig. \ref{phase_active_FB} of appendix \ref{sec:Fz_balance}, we show the behavior of vesicles where we apply an external force that balances the total vertical component of the active forces. For a living cell, this condition corresponds to assuming that actin filaments that are pushing the top membrane upwards exert an equal and opposite force on the bottom membrane. We see that except at the lowest $E_{ad}$, there is no qualitative difference, compared to the previous results (Fig.\ref{phase_active}).

The importance of coupling the force to curvature, is demonstrated by simulating the adhesion of a vesicle with flat active proteins (zero spontaneous curvature, $c_0=0$). As in previous studies \cite{miha2019,graziano2019cell}, we find the formation of long protrusions, that are highly dynamic (Fig. \ref{active_c0_0} of appendix \ref{sec:active_c0_0}). Due to the adhesion, the long protrusions are found to often grow along the substrate. However, when they point upwards, they lead to partial detachment of the vesicle. Clearly, active forces that are not coupled to curvature do not contribute to effective spreading and adhesion.

When comparing our results with experimental observations of the shapes of adhered cells, we begin by noting that cells undergo a much diminished spreading (or strong retraction) when actin polymerization is inhibited (adhered area decreases by factor of $\sim4$ \cite{Sheetz2008}). This suggests that the bare adhesion of the cell to external substrates is typically low, so that in terms of our model cells are usually in the regime of low $E_{ad}$. In this regime, we demonstrate that self-organization of the actin polymerization recruited by curved membrane proteins can increase the adhered area by factors that are similar to those observed experimentally (Fig.\ref{ad_area_active}b,c). 

However, the actin polymerization in the cell does more than just provide a protrusive force, as we assumed in our model. The actin retrograde flow produces shearing forces that triggers the growth of integrin-based adhesion complexes \cite{cavalcanti2007cell,schaufler2016selective}. This suggests that the activity of actin polymerization also effectively increases $E_{ad}$ for the cell, compared to the actin-inhibited cell. Similarly, increased adhesion strength ($E_{ad}$) allows for stronger mechanical coupling between the actin filaments and the substrate, inducing a larger effective protrusive force $F$ \cite{chan2008traction,gong2018matching}. These effects mean that when comparing our model to cell shapes, the effective actin protrusive force $F$ and the effective value of $E_{ad}$ are not independent of each other. 

Many adhered cells are found not to be circularly spread, but have a distinct spindle-like shape with usually two oppositely formed lamellipodia protrusions. This typical shape appears naturally in our model when the density of the curved proteins is below the pancake transition value, and the adhered vesicle assumes the elongated two-arc shape (Figs. \ref{phase_active}, \ref{phase_active_pancake}). Note that since the critical density for the pancake transition increases for decreasing $E_{ad}$ (Fig.\ref{phase_active_pancake}b), we expect that cells can transform from the pancake to the elongated two-arc shape with decreasing adhesion strength. This is indeed observed in experiments \cite{cavalcanti2007cell,schaufler2016selective,singh2020cell}. The morphology of two oppositely oriented lamelipodia (similar to our two-arc shapes) was observed to stretch cells, and is sometimes utilized to drive cell division \cite{kee2012mechanosensory,flemming2020cortical}.

\subsection{Spreading dynamics}
\begin{figure}[ht]
\centering
\includegraphics[scale=0.9]{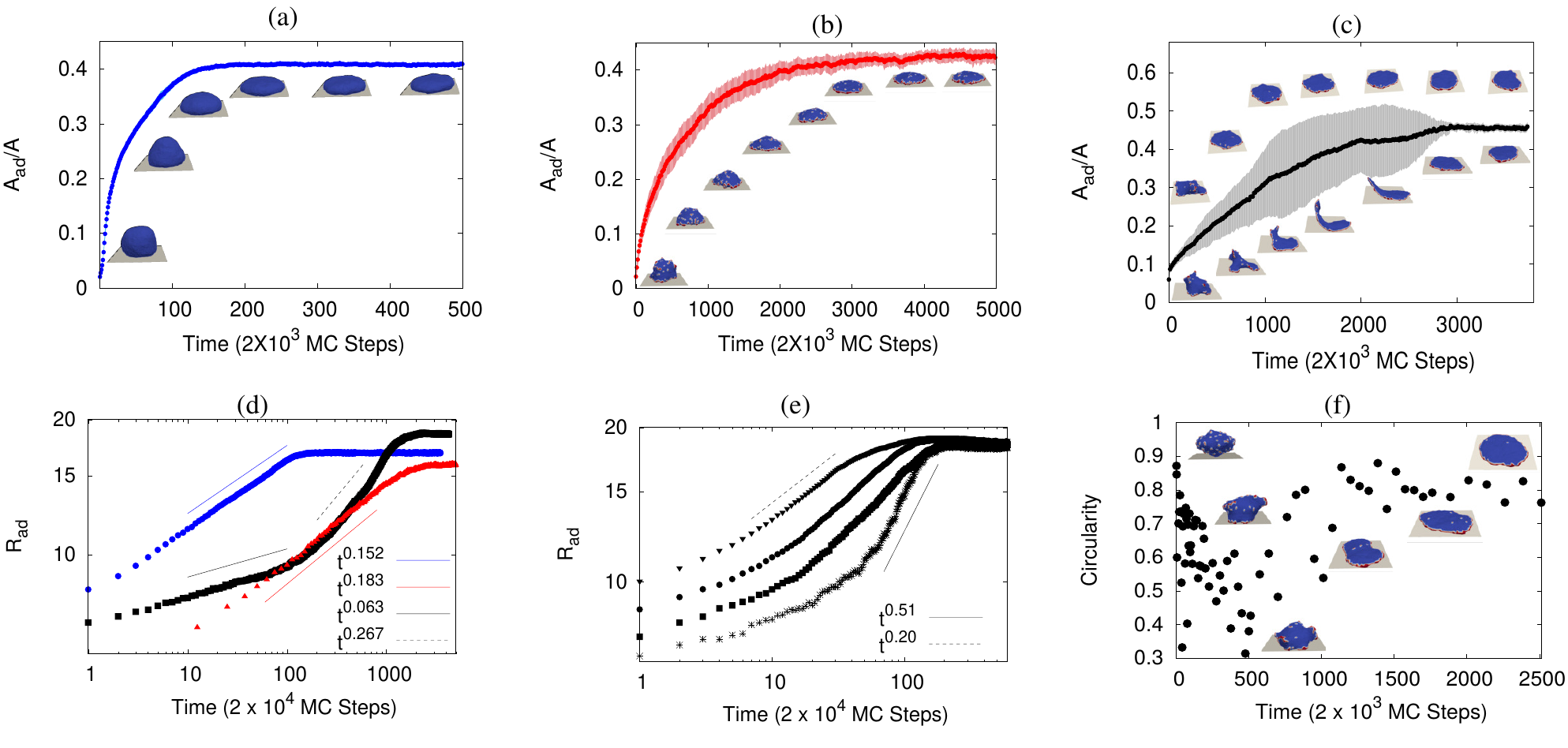}
\caption{Dynamics of the spreading process. (a) Variation of adhered area with time for a protein free vesicle. Here, we use $E_{ad}=5.0 ~k_B T$. (b) Variation of adhered area with time for a passive vesicle with $\rho=13.82 ~\%$ and $E_{ad}=0.50 ~k_B T$. (c)  Variation of adhered area with time for an active vesicle with $F=4 ~k_B T/l_{min}$, $\rho=10.36~\%$ and $E_{ad}=0.50~ k_B T$. We note that the protein free vesicle adhere much faster than the other cases. The spreading of passive vesicle is very smooth, while the active vesicle spreading is very noisy. Here, the unit of time (t) is $2 \times 10^3$ MC steps for (a)-(c). (d) Variation of adhered radius ($R_{ad}$) with time, for protein-free, passive and active vesicles. The blue circle is for protein-free vesicle, the red triangles are for passive vesicle and the green boxes are for active vesicle. We note that the growth of a protein free vesicle and passive vesicle is uniform in the beginning, and then saturates, however, for active case, there are two phases of growth in the small time regime before $R_{ad}$ saturates. For protein free vesicle, $R_{ad}(t) \sim t^{0.152}$, while for a passive vesicle, $R_{ad}(t) \sim t^{0.183}$. Unlike the protein-free and passive vesicles, the active vesicle exhibits two growth phases: a slower initial growth of  $R_{ad}(t) \sim t^{0.063}$ followed by a faster growth $R_{ad}(t) \sim t^{0.267}$. Here, we use $E_{ad}=5.0 ~k_B T$ for protein-free vesicle, and $\rho=10.36 ~\%$ and $E_{ad}=0.50 ~k_B T$ for passive and active vesicles. For active vesicle, we use $F=4 ~k_B T/l_{min}$. (e) Variation of $R_{ad}$ with time for an active vesicle with various values of $E_{ad}$: from bottom to top, star symbols are for $E_{ad}=0.25 ~k_B T$, boxes are for $E_{ad}=0.50$, circles are for $E_{ad}=1.0 k_B T$ and triangles are for $E_{ad}=2.0 ~k_B T$. Note that the fast growth phase takes longer to appear for lower $E_{ad}$, while the growth exponent of the fast growth phase increases with $E_{ad}$. We vary $E_{ad}$ from $0.25 ~k_B T$ to $2.0~ k_B T$, that gives rise to variation in the exponent from $0.51$ to $0.20$. Here we use $\rho=10.36 ~\%$ and $F=4~ k_B T/l_{min}$. The unit of time (t) is $2 \times 10^4$ MC steps for (d)-(e). (f) The circularity of the vesicle shows non-monotonic variation with time. Here, we use $F=4  ~k_B T/l_{min}$, $E_{ad}=0.50 ~k_B T$ and $\rho=10.36 ~\%$.}
\label{t_vs_A} 
\end{figure}

We compare our results for the spreading dynamics of a protein-free vesicle, a vesicle with passive-curved and active-curved proteins respectively, in Fig. \ref{t_vs_A}(a-c). The vesicles with proteins are shown in the interesting regime of low $E_{ad}$. We note that the active vesicle spreading is much noisier than the passive spreading. This is because in the active case the vesicle may transiently get locally de-adhered from the substrate, which gives rise to large variations in the measurement of the AAF (Fig. \ref{t_vs_A}(c)). In Fig. \ref{spreading_dynamics} of appendix \ref{sec:spreading_dynamics}, we plot the cross-sectional shapes, side-views and three-dimensional shapes, of the spreading vesicles as function of time, for all three cases. Clearly, the passive systems are observed to spread more isotropically, compared to the active system. The anisotropic spreading of the active vesicles is quantified in Fig. \ref{t_vs_A} (f), where we see a sharp reduction in the circularity of the adhered region during the initial stages of spreading.

Next, we  plot the increase of the adhered radius ($R_{ad}$) as function of time, which is defined as $\sqrt{A_{ad}/\pi}$, where $A_{ad}$ is the adhered area (Fig. \ref{t_vs_A}(d-e)). 
We find that for the passive systems the adhered radius ($R_{ad}$) grows with time as $\sim t^\beta$, where the exponent $\beta$ is different for the different cases. Although this plot is given in MC time-steps, and does not include the hydrodynamic effects of the the membrane flow and the fluid flow within and around the vesicle, the calculated dynamics of the passive vesicles resemble the experimental observations for spreading artificial vesicles \cite{streicher2009integrin}.

For the active case (Fig. \ref{t_vs_A}e), we averaged over those cases which adhered smoothly without significant events of de-adhesion, in order to get a less noisy curve. For this case, we find that there is a slower growth in the beginning (Fig. \ref{t_vs_A}d), followed by a faster growth regime. The exponent of the faster growth stage depends on $E_{ad}$ (Fig. \ref{t_vs_A} (e)). The slower initial growth is due to the low circularity of the active vesicle, with the protein aggregates spread over the vesicle and pushing the membrane upwards and in uncoordinated manner. Once the proteins form the circular aggregate along the contact-line, they induce a very efficient and rapid spreading, which corresponds to the fast growth phase.

The calculated spreading dynamics for an active vesicle resemble several aspects experimentally observed in spreading living cells \cite{Sheetz2004PRL,dubin2004nanometer,Sheetz2014}: (i) The active vesicles exhibit an initially accelerating radial growth, followed by a growth with almost constant velocity (Fig. \ref{t_vs_A} (c,d)). These features are observed in living cells, and do not appear for our passive vesicles. (ii) The initiation of the rapid spreading phase takes longer to appear for cells on substrates of lower adhesiveness, as we also find (Fig. \ref{t_vs_A} (e)). (iii) Compared to the passive system, the active vesicle initially grows more slowly (Fig. \ref{t_vs_A} (d)), similar to experimental observations \cite{cuvelier2007universal}.

Other features of cell spreading are also manifested in our spreading active vesicles: Similar to the case of spreading cells \cite{Sheetz2010}, we find that the circularity of the spreading active vesicles decreases sharply during the beginning of the spreading process, and recovers slowly afterwards.
Furthermore, the active forces often give rise to the transient upwards detachment of the leading edge in our simulations (Fig.\ref{t_vs_A} (c)), resembling the ruffles observed at the leading edge of spreading cells \cite{Sheetz2011,Nils2017,begemann2019mechanochemical,simeonov2019high}.

Note that since we do not conserve the vesicle volume, we find that the volume strongly decreases as the vesicles spread. We discuss this in more detail in the appendix \ref{sec:volume} (Fig. \ref{area_volume}).

\subsection{Motile vesicles at low protein densities}
\begin{figure}[ht!]
\centering
\includegraphics[scale=0.8]{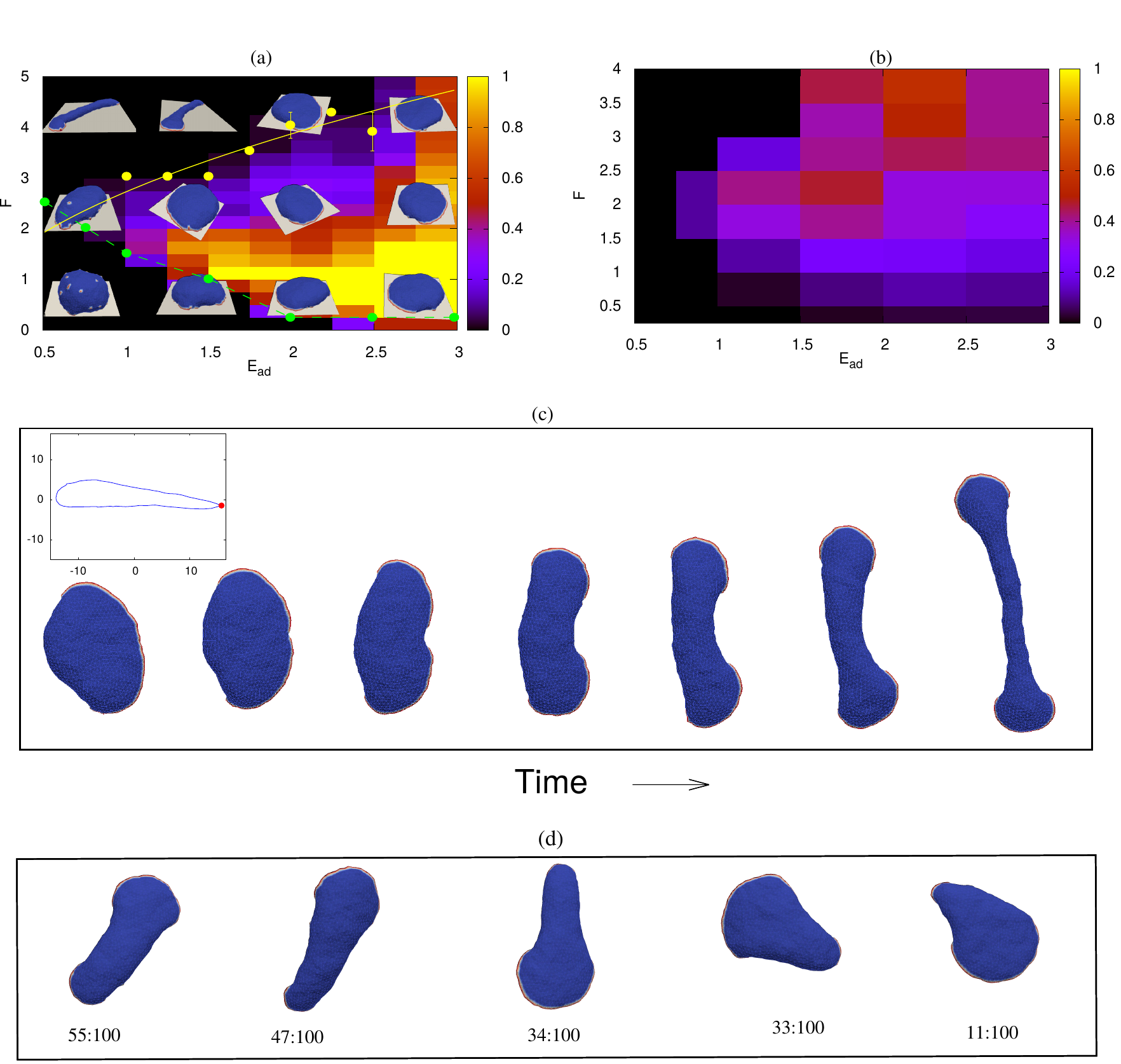}
\caption{Crescent-shaped (motile) vesicle, its speed and stability. (a) Probability for a vesicle to be found in a crescent-shaped state (background color), in the $E_{ad}-F$ plane for small $\rho=3.45~\%$. In the region where there is non-zero probability of obtaining a crescent-shaped vesicle, the  snapshots of a crescent-shaped vesicles are shown, otherwise a typical snapshot is shown. The yellow solid line is the analytical prediction separating the two-arc phase to a crescent-phase (Eq.\ref{crescentEad}). Here, we fit the numerical data and obtain a good fit at $ E_{ad} \simeq 2.04 ~F^2$. We also numerically calculated the prefactor and it turns out to be $\simeq 1.96$, close to the value obtained by fitting the curve (see Fig. \ref{delta_A} for details). The yellow dots are the simulation point separating the two-arc shapes from crescent-shaped vesicle: For a given $E_{ad}$, these dots represent the value of $F$, above which we do not find crescent shaped vesicles. The green dashed line separates the region of vesicles with disordered small protein clusters (below) and the regime of clustered proteins (either two-arc or crescent shape). Here, we estimate this line by measuring the largest cluster, and if the largest cluster is less than $30 \%$ of the total proteins, we regard it to be in the disordered state of small protein clusters. The snapshots are shown for $E_{ad}=$ 0.50, 1.5, 2 and 3 (in units of $k_B T$) and $F=$ 0.50, 2 and 4 (in units of $k_B T/l_{min}$). (b) Speed of the crescent-shaped vesicle, scaled by $E_{ad}$. We calculate the speed as the displacement of COM per MC step (divided by $E_{ad}$), and then normalize all the values by the maximum speed found for the range of parameters shown here, where the maximum value of the speed is $0.164 ~ l_{min}$ per MC step. For large $E_{ad}$ but small $F$, the crescent vesicles exhibit diffusive behaviour (Fig.\ref{MSD}). (c)  Spontaneous transition from crescent-shape to two-arc shape, demonstrating the transient nature of the motile shapes. In the inset, we show the cross-sectional view of a motile crescent-shaped vesicles. The red dot is showing the location of the high protein density. (d) Examples of asymmetric two-arc shapes, which exhibit weak residual motility. Below each shape, we also mention the ratio of smallest to largest cluster. For both (c,d) we use $E_{ad}=3.0 ~k_B T$, $\rho=3.45 ~\%$ and $F=4 ~k_B T/l_{min}$.}
\label{phase_crescent} 
\end{figure}

In the regime of low (curved and active) protein density, where there are not enough proteins to form a circular aggregate around the cell rim (pancake shape), we find that the vesicles can form motile shapes. By "motile" we mean that the active proteins form a single large aggregate on one side of the vesicle, that results in an unbalanced force that pushes the vesicle along the adhesive substrate. Such motile crescent shapes are shown for example in Figs. \ref{phase_crescent} (a,c). 

In Fig. \ref{phase_crescent}a we show the regime of active force and adhesion energy that gives rise to the motile crescent shapes, at very low protein density ($\rho=3.45~\%$). We find that the regime where the crescent shapes appear coexists with two-arc shapes, and the two can transiently convert into each other (see for example Fig. \ref{phase_crescent}(c)). When moving, the crescent shape remains persistent by maintaining a sharp leading edge (Fig. \ref{phase_crescent}(c), inset), due to the active force concentrated in a single cluster, while the rear region is less curved due to minimization of bending energy and area conservation. 

In Fig. \ref{phase_crescent}(b) we plot the average velocity of the crescent shapes, calculated from the simulations, divided by an effective friction coefficient that is assumed to be linear in $E_{ad}$ (the vesicle speed before this scaling is shown in Fig. \ref{crescent_speed} of appendix \ref{sec:crescent_speed}). Note that also the two-arc shapes are weakly motile, as the protein aggregates at each end of the cell are not identical in size and there is a small residual force (Fig. \ref{phase_crescent}(d)). The crescent shapes exhibited persistent motility with a well defined velocity in the high force regime, while for weak active forces they exhibited diffusive motility (see Fig. \ref{MSD} of appendix \ref{sec:MSD}).

The region of crescent shapes in Fig. \ref{phase_crescent}(a) is bounded by two transition lines. The lower line denotes the line below which the proteins do not form large clusters. In this regime the proteins form disordered small clusters, and the vesicle remains approximately hemispherical. Above this line, the proteins form one or two large aggregates, thereby enabling the formation of the crescent or two-arc shapes. Above the transition line the large adhesion energy or strong active forces induce the sufficiently high curvature at the vesicle contact line, which concentrates the proteins and drives the formation of large aggregates. This is similar to the pancake (and two-arc) transition line denoted in Fig. \ref{phase_F_rho}, where increasing $E_{ad}$ corresponds to higher $\rho$ at the contact-line region. Note that the direct protein-protein interaction strength $w$ plays a minor role in this transition, which can occur even for $w=0 ~k_B T$ (Fig. \ref{different_w}(a) of appendix \ref{sec:various_w}).

The upper transition line that bounds the crescent shapes regime in Fig. \ref{phase_crescent}(a) can be estimated by comparing the bending and adhesion energies of a two-arc shape versus the crescent shape. Compared to the crescent shape, the two-arc shape has a lower adhesion energy, since the elongated cylindrical part is more weakly adhered as it is devoid of curved proteins. It is also more strongly curved compared to the circular shape of the crescent vesicle. On the other hand, the work done by the active forces that elongate the two-arc vesicle counts as a negative energy contribution. The net difference between the two classes of shapes can be approximately written as
\begin{equation}
    \Delta F=-NFL+2\pi RL\frac{\kappa}{2R^2}+E_{ad}\Delta A
    \label{dFcrescent}
\end{equation}
where $R$,$L$ are the radius and length of the cylindrical segment of the two-arc shape, $N$ is the number of proteins that pull the membrane at the two ends of the cell and $\Delta A$ is the difference in adhered area between the two-arc and crescent shape. We first minimize with respect to the radius $R$, taking the total area of the cylindrical segment to be conserved $A_{cyl}=2\pi RL$. We find that $R=2\pi\kappa/(NF)$. We substitute this value in Eq. \ref{dFcrescent}, and calculate the critical adhesion energy at which $\Delta F=0$
\begin{equation}
    \Delta F=0\rightarrow E_{ad}=\frac{A_{cyl}}{2\kappa\Delta A}\left(\frac{NF}{2\pi}\right)^2
    \label{crescentEad}
\end{equation}
We assumed here for simplicity that $\Delta A$ is constant along this transition line, which is approximately obeyed by the simulation results (Fig. \ref{delta_A}, appendix \ref{sec:delta_A}). The relation in Eq. \ref{crescentEad} appears to capture correctly the essence of the transition between the two shapes, as shown in Fig. \ref{phase_crescent}a. We fit the Eq. \ref{crescentEad} with simulation data points, and obtain the relation $E_{ad} \simeq 2.04~F^2$, that matches well with the simulation. We also estimate this prefactor of $F^2$ by calculation $\Delta A$ and $A_{cyl}$ from simulation, and obtain a value $\simeq 1.96$ close to the value obtained by fitting the curve (see \ref{delta_A} for details).

When comparing the motile shapes that our model produces, they bear similarity to the shapes of different motile cells \cite{mogilner2009shape}. Similar to experimental observations \cite{dimilla1991mathematical,klank2017biphasic,liu2020cell}, we predict a maximum of the migration speed as function of the adhesion strength (Fig. \ref{phase_crescent}b). These results indicate that the coupling of curved proteins that recruit the actin polymerization, and adhesion, can self-organize into a spontaneously motile shape. However, these ingredients give rise to rather delicate and transient motility, which can either disappear spontaneously (Fig. \ref{phase_crescent}(c)) or when confronted with an external perturbation (Fig. \ref{barrier}, appendix \ref{sec:barrier}). Within the wider context of active-matter systems, our motile vesicles can be compared to recent works that have shown similar symmetry breaking that is driven by self-organization of active elements \cite{abaurrea2019vesicles}.

\section{Discussions}
\label{sec:sum}

We have shown here how interacting curved membrane proteins, passive and active, affect the process of vesicle spreading and adhesion. We find that large density of passive curved proteins can greatly enhance the adhesion of vesicles on low adhesion substrates. Coupling the curved proteins with active protrusive forces extends this enhancement to lower densities of curved proteins. By spontaneously self-organizing curved proteins at the cell-substrate contact line, the active forces drive a shape transition into a flat geometry with high adhered area and robust spreading. At very low densities of curved proteins the protrusive activity can stabilize either spindle-like elongated cells, or motile crescent shapes.

Our simplified model does not contain all the complexities of a real cell, which strongly affect its final adhered shape. One such component, the network of stress-fibers, is known to determine the cell shape in many cell types \cite{schwarz2013physics}. In addition, the cytoskeleton and internal organs (such as the nucleus) hinder the shape changes of the cell, and exert volume constraints. Future extensions of our model can include additional components of the cell adhesion process. For example, we could add non-uniform adhesion that is activated closer in proximity to the curved proteins, to describe the activation of adhesion by actin retrograde flow \cite{bershadsky2003adhesion,geiger2009environmental,gardel2010mechanical,Ibata2020nascent}. Nevertheless, our model describes many features of spreading cells, allowing to relate the observed cell spreading dynamics and the cell shape to the parameters of the model. 

Observations in living cells emphasize the central role played by actin polymerization during cell spreading and adhesion. These observations suggest that in living cells the membrane density of highly curved proteins is relatively low, and cells are not likely to be in the regime where a high density of curved proteins alone drives the spreading and adhesion (Fig. \ref{phase_passive}). Loading the membrane with a large density of such curved proteins may be problematic for the cell, and limit its ability to dynamically control and modify its spreading and adhesion strength. Our model demonstrates that by having a low bare adhesion, and low density of curved proteins, the cell can achieve robust and dynamic adhesion by activating the protrusive force of actin polymerization, in a highly localized and self-organized pattern. The spontaneous aggregation of the curved proteins along the cell-substrate contact line, driven by the actin-induced forces (and attractive direct interactions between the proteins), provides a highly controllable mechanism for cell spreading and adhesion.

In addition to non-motile steady-state shapes of adhered vesicles, we found that in the low $\rho$ regime the vesicles may form a polarized, crescent shape, that is motile (Fig.\ref{phase_crescent}). This motile vesicle resembles the shapes of motile cells, that depend on adhesion \cite{barnhart2017adhesion}, and demonstrates that the combination of curved proteins that recruit the actin polymerization, and adhesion, provide a minimal set of ingredients needed for motility. However, in order to make the polarization that drives the motility robust and persistent (as opposed to transient), cells have evolved additional biochemical feedbacks of various types \cite{ridley2003cell,maiuri2015actin,rappel2017mechanisms}. Our model does not contain many components that play important roles in cell motility, such as contractility, and more realistic treatment of the actin-adhesion coupling, such as catch and slip-bond dynamics. Our results however highlight that curvature-force coupling, with adhesion, provide the basic coarse-grained components that can self-organize to spontaneously break the symmetry and form a motile system.
\section{Acknowledgements}
We thank Orion Weiner, Benjamin Geiger, Ronen Zaidel-Bar, Robert Insall, Sam Safran, Jeel Raval and Wojciech Gozdz for useful discussions. N.S.G. acknowledges that this work is made possible through the historic generosity of the Perlman family. N.S.G. is the incumbent of the Lee and William Abramowitz Professorial Chair of Biophysics and this research was supported by the Israel Science Foundation (Grant No.1459/17). A.I. and S.P. acknowledge the support from Slovenian Research Agency (ARRS) through program No. P2-0232 and the funding from the European Union's Horizon 2020 - Research and Innovation Framework Programme under grant agreement No. 801338 (VES4US project).

\appendix
\renewcommand{\theequation}{A-\arabic{equation}}
\setcounter{equation}{0}
\renewcommand{\thefigure}{A-\arabic{figure}}
\setcounter{figure}{0}
\section{Simulation details}
\label{sec:simulation_detail}
The time evolution of the vesicle in our MC simulations consists of \cite{miha2019}: (1) vertex movement, and (2) bond flip. In the vertex movement, a vertex is randomly chosen and attempt to move by a random length and direction within a sphere of radius $s$ drawn around the vertex. In the bond flip movement, a single bond is chosen, which is a common side of two neighbouring triangles.  The bond connecting the two vertices in diagonal direction is cut and reestablished between the other two, previously unconnected vertices. In order to satisfy self avoidance, the ratio of maximum and minimum bond length, i.e., $l_{max}/l_{min}=1.7$ and the maximum possible displacement of a vertex in a given attempt is taken to be $s=0.15$ in units of $l_{min}$.
 
We use Metropolis algorithm to update our system. Any movement that increases the energy of the system by an amount $\Delta E$ occurs with rate $exp(- \Delta E/k_BT)$, otherwise if the movement decreases the system energy, it occurs with rate unity. We let the system evolve according the above rule and wait till the system reaches steady state. All the average quantities are measured after the system reaches steady state. 

In the simulations presented in this paper we use the following model parameters: Total number of vertices, $N=1447$, the bending rigidity $\kappa =20 ~k_B T$, the protein-protein attraction strength $w=1 ~k_BT$. The width of the potential, $\Delta z$ is taken to be $l_{min}$. Among all the $N$ vertices, $N_c$ of them are occupied by curved membrane proteins. The spontaneous curvature at all $N_c$ vertices are taken to be $c_0=1~ l^{-1}_{min}$, unless stated otherwise. We chose this set of parameters to be in the interesting regime where the curved proteins form aggregates, and exhibit a force-driven phase-separation into a pancake-like shape \cite{miha2019}.
\section{Comparison of simulated and detailed numerical solutions of the adhered vesicle shape}
\label{sec:jeel}
\begin{figure}[ht]
\centering
\includegraphics[scale=1.5]{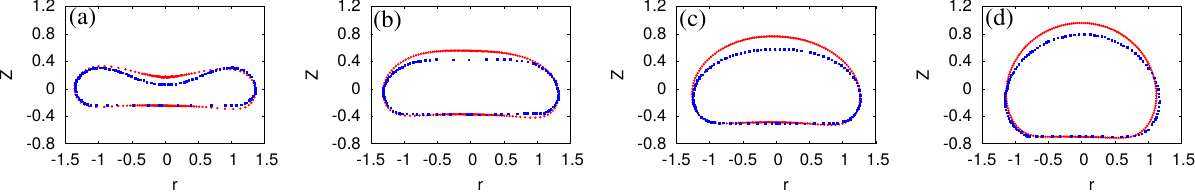}
\caption{Comparison of the results of MC simulation with detailed numerical solutions \cite{raval2020shape,bibissidis2020}. The red circles are for simulation results and the blue boxes are for detailed numerical solution. (a) $\tilde{w}=6.4$, $v=0.545$, (b) $\tilde{w}=6.4$, $v=0.75$, (c) $\tilde{w}=6.4$, $v=0.85$ and (d) $\tilde{w}=6.4$, $v=0.95$. For detailed numerical solution, the data is extracted from Fig. 5 of ref. \cite{bibissidis2020} using the `digitize image' tool from `OriginLab' software.}
\label{compare_jeel} 
\end{figure}

In this section, we compare the results of our MC simulations for the shapes of adhered protein-free vesicles, with detailed numerical solutions that appeared recently in Ref. \cite{raval2020shape,bibissidis2020}. In the detailed numerical solutions, the parameters used are $\tilde{w}$, the scaled adhesion strength and $v$, the reduced volume. The parameter $\tilde{w}$ is defined as, $\tilde{w}=E_{ad} R_s^2/\kappa$, where $E_{ad}$ is the adhesion energy per unit adhered area and $R_s$ is the radius of a spherical vesicle with same volume as the original. Since, in our model we define $E_{ad}$ as the adhesion energy per vertex, we properly scaled it before comparison. In \cite{raval2020shape,bibissidis2020}, the reduced volume $v$ is fixed, however, in our model, we can not fix $v$ before the adhesion and spreading dynamics. In order to access different values of $v$ for the same $\tilde{w}$, we use the osmotic pressure difference $p = p_{inside} - p_{outside}$, that adds one more energy term $-p V$ to Eq. \ref{eq:energy}, where $V$ is the total volume of the vesicle \cite{miha2019}. In Fig. \ref{compare_jeel}, we show the comparisons of the shapes of the vesicle from our simulation and the detailed numerical solution (as given in Fig. 5 of ref. \cite{bibissidis2020}). We note that the shape is comparable to our MC simulations, and the agreement is very reasonable, thereby validating the MC approach. 

\section{Analytical model}
\label{sec:analytical}
We now present an analytic calculation of the adhered shape of the vesicle, in the presence of curved proteins and active forces. This calculation correctly describes the qualitative features that we found in the simulations, but fails quantitatively. The main failure is the use of protein-protein interactions that are good for the dilute limit (Eq.\ref{rhoTerm}), and do not capture the large density increase at the contact line as in the simulations. In regimes where the protein-protein interactions are not playing an important role, for example when the active force is large, we find quantitative agreement (Figs.\ref{compare_active},\ref{active_analytical} below).

\begin{figure}[ht]
\centering
\includegraphics[scale=0.7]{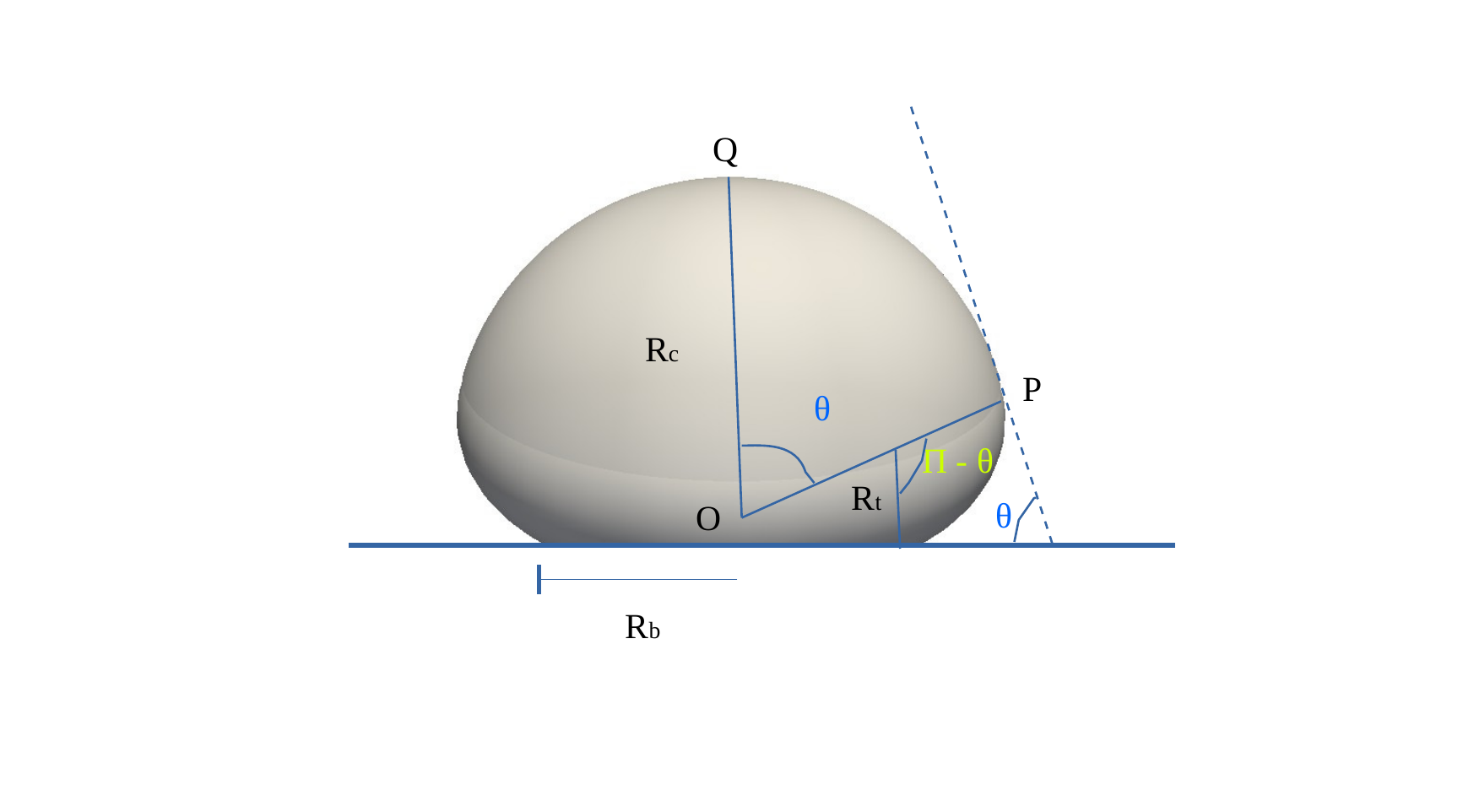}
\caption{Schematic representation of the analytical model. $R_c$ is the radius of the spherical cap, $R_t$ be the radius of the annulus part (torus), $R_b$ be the radius of the circular disk at the base. $\theta$ is the angle between  the line perpendicular to the plane `OQ' and the line joining the center of sphere `O' and the point `P', where the spherical cap and torus section meets.}
\label{schematic}
\end{figure}

We assume the average shape of the vesicle consists of three parts: (1) The base area, which is having a circular shape, with radius $R_b$, (2) the annulus part curving around the circular base, which is the part of a torus, with radius of the tube $R_t$, and (3) the spherical cap, which is a part of sphere, with radius $R_c$ (see Fig. \ref{schematic}). $\theta$ is the angle between the vertical line passing through the centre of the sphere (OQ), and the line joining the centre of sphere and the contact point of sphere and the torus (OP), as shown in Fig. \ref{schematic}. We neglect any thermal fluctuation in the shape of the vesicle. Total area of the vesicle, $A= A_b + A_t + A_c \simeq \pi R_b^2 + 2 \pi (1-cos \theta) R_c^2 + 2 \pi R_t R_b (\pi - \theta ) $, where $A_b$ denotes the area of the base, $A_t$ is the area of the annulus part (torus) and $A_c$ is the area of the spherical cap. Total area $A$ is taken to be constant \cite{lipowski2005}. 

Let $\rho $ be the density of the spontaneous curvature $c_0$, $\rho_c$ is the density on cap, and $\rho_t $ is the density on the annulus, such that, $\rho A = \rho_t A_t + \rho_c Ac$. Here, we assume that bottom part of the vesicle does not contain proteins, which is also seen in the simulations. Now, the adhesion energy is given by, 

\begin{equation}
    W_A= - E_{ad} \pi R_b^2,
\end{equation}

the bending energy is given by, 
\begin{equation}
W_b = \frac{1}{2} \kappa A_t ( \frac{1}{R_b} + \frac{1}{R_t} -c_0 \rho_t )^2 + \frac{1}{2} \kappa A_c (\frac{2}{R_c} - c_0 \rho_c)^2,
\end{equation}

the protein-protein nearest neighbour attraction energy is of the form,
\begin{equation}
W_d = -\frac{w}{2} (A_c \rho_c^2 + A_t \rho_t^2),
\label{rhoTerm}
\end{equation}

the energy due to active force,
\begin{equation}
W_F= - F R_t \rho_c A_c - F R_b \rho_t A_t,
\end{equation}

and finally, the entropy, 
 
\begin{equation}
    S=-A_c \{ \rho_c ln (\rho_c) + (1- \rho_c) ln (1-\rho_c)\}  -A_t \{ \rho_t ln (\rho_t) + (1- \rho_t) ln (1-\rho_t)\},
\end{equation}

where, $E_{ad}$ is the adhesion  energy per unit adhered area and $\kappa$ is the bending rigidity. Note that the entropy is only due to the thermal fluctuation of proteins and any entropy due to the thermal fluctuation in the vesicle shape is  not considered here. In the calculation of entropy, we assume that only the spherical cap of the vesicle and the annulus part (torus part) contain proteins while the base of the vesicle does not contain proteins, which is also seen in simulations. Since, the slope of the surface changes continuously along the angle $\theta$, the slope at the contact point of torus and spherical cap should be the same. This gives us a constrains, $R_c sin \theta = R_b + R_t sin \theta$.
The free energy is given by,

\begin{equation}
F=W_b + W_A + W_d + W_F- T S
\label{eq:F_analytical}
\end{equation}

We assume the total area $A$ to be a constant. We then minimize the free energy (Eq. \ref{eq:F_analytical}), in the ($R_b, \rho_t, \theta$) plane, and express other parameters in terms of these three parameters. More explicitly, we solve the equations, $\frac {\partial F}{\partial R_b} =0$, $\frac {\partial F}{\partial \rho_t} =0$, and $\frac {\partial F}{\partial \theta} =0$ with the constraint $\frac {\partial^2 F}{\partial R_b^2} > 0$ and $\frac {\partial^2 F}{\partial \rho_t^2} > 0$, $\frac {\partial^2 F}{\partial \theta^2} > 0$. Among several solutions, we consider the physical one.

In the analytical model, we define $E_{ad}$ as the adhesion energy per unit of adhered area (having dimension of energy/length$^2$), while, in the simulation, we define it as the adhesion energy per adhered number of vertex (with dimension of energy). In order to compare the simulation and analytical results, we properly scale $E_{ad}$ such that the definition becomes consistent in both the cases. The value of the parameters used here are: $A=2200~ l_{min}^2$, which is approximately the average area of a unstretched  vesicle in our simulation, $c_0=1.0 ~l^{-1}_{min}$, $\kappa=20 ~k_B T$, $w=1.0 ~k_B T$. The value of $k_B T$ is taken to be unity.

\begin{figure}[ht]
\centering
\includegraphics[scale=0.75]{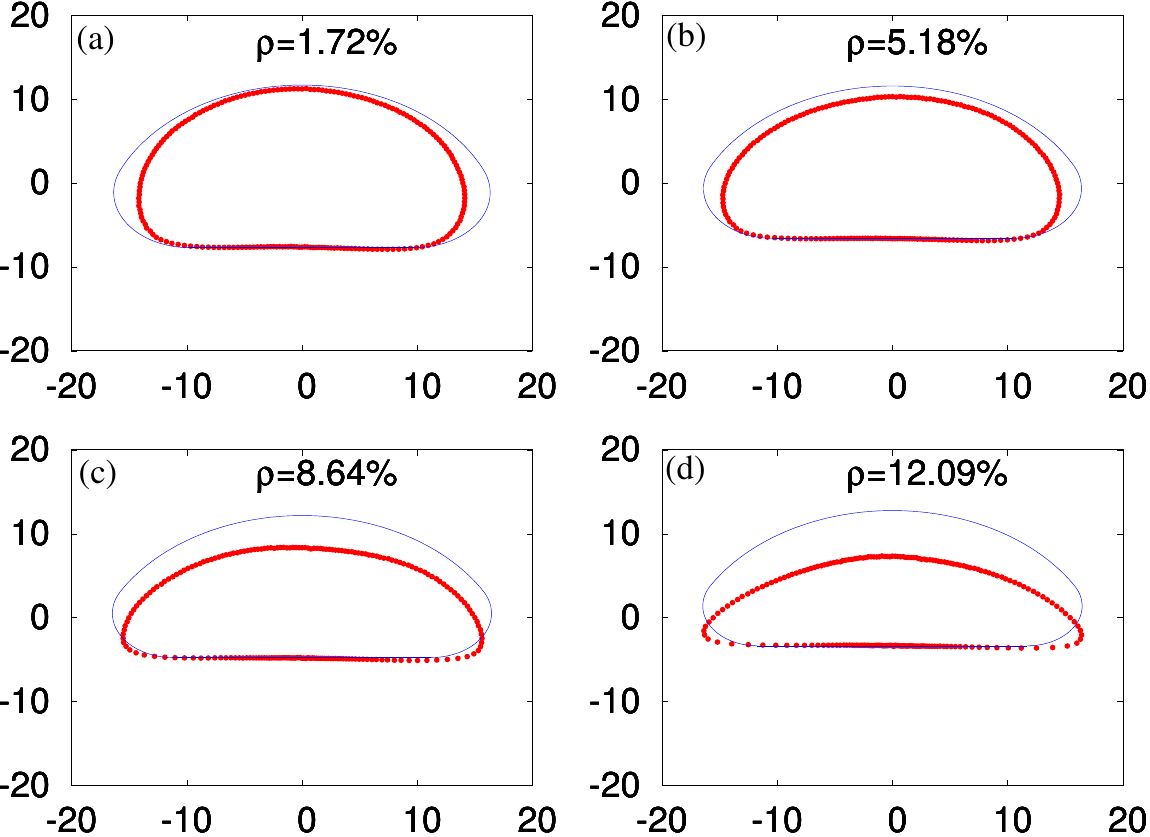}
\caption{Comparison of the shape of vesicle for simulation results and analytical results for passive case with $E_{ad}=0.50~ k_B T/l_{min}^2$ and various $\rho$. Red circles are for simulation results and blue line is for analytical results.}
\label{compare_passive} 
\end{figure}

\begin{figure}[ht]
\centering
\includegraphics[scale=1.5]{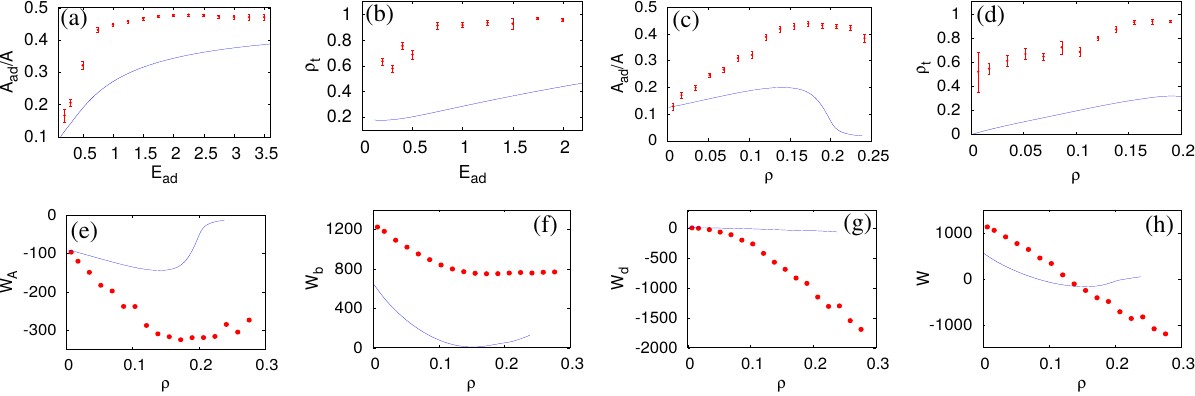}
\caption{Results for analytical model for passive case ($F=0$) and comparison with simulation. Red circles are for simulation results and blue line is for analytical results. (a)  Fraction of adhered area with adhesion strength $E_{ad}$ for $\rho=10.36 \%$. The adhered area increases and tends to $1/2$ for large $E_{ad}$. (b) The density of curved proteins in the annulus part of the vesicle ($\rho_t$) as a function of $E_{ad}$ for $\rho=10.36 \%$. It shows that the density $\rho_t$ increases and tends to unity for large $E_{ad}$. Here also, the analytical prediction is smaller that the simulation results. (c) Fraction of adhered area with the protein density $\rho$ for a given $E_{ad}$. We note that the analytical prediction also shows non-monotonic variation as seen in the simulation.  (d) $\rho_t$ with $\rho$ for given $E_{ad}$. (e) The adhesion energy ($W_A$)) as a function of $\rho$. (f) The bending energy ($W_b$) as a function of $\rho$. (g) The protein-protein interaction energy ($W_d$) as a function of $\rho$. (h) The total energy ($W$) as a function of $\rho$. For (c) to (h), we use $E_{ad}=0.50 ~k_B T/l_{min}^2$. For analytical results, we use here $A=2200~l^2_{min}$, which is the average area of a unstretched  vesicle in our simulation, $c_0=1~l^{-1}_{min}$, $\kappa=20 ~k_B T$, $w=1.0~ k_B T$. For analytical results,  $k_B T$ is taken to be unity. For MC simulations, the other parameters are same as Fig. \ref{phase_passive}.}
\label{passive_analytical} 
\end{figure}

For passive case ($F=0$), we compare our analytical results for the cross-sectional shape of the vesicles with different $\rho$ in Fig. \ref{compare_passive}. We note that the effect of increasing $\rho$ is not very strong in the analytical model. We show the comparison of adhered fraction, density of proteins in the curved regions ($\rho_t$) and the different energies in Fig. \ref{passive_analytical}. We note that the fraction of adhered area increases with $E_{ad}$ similar to our simulations. With increasing $E_{ad}$ the vesicle becomes more and more flat and tends to the value $1/2$ for large enough $E_{ad}$ (Fig. \ref{passive_analytical}(a)). The analytical prediction is however lower than the simulation results.  We also measure the density of curved proteins in the annulus part, $\rho_t$. As $E_{ad}$ increases, $\rho_t$ also increases and tends to unity for simulations, however, the analytical prediction is very low (Fig. \ref{passive_analytical}(b)). This indicates that for large $E_{ad}$ most of the proteins are aggregated in the curved annulus region. In our simulation, we also note that for large $E_{ad}$, we do have a pancake-like phase, where all the proteins are clustered in the curved region. We also measure the fraction $A_{ad}/A$ as a function  of $\rho$ for given $E_{ad}$ (Fig. \ref{passive_analytical}(c)). We note that similar to the simulation results, our analytical model also shows non-monotonic variation in $A_{ad}/A$ with $\rho$, however, here also, the quantitative comparison is not very well. Finally, we also show the variation of $\rho_t$ with $\rho$, which also show an increase similar to simulation results (Fig. \ref{passive_analytical}(d)). 

We also compare the different energies in the lower panel of Fig. \ref{passive_analytical} for simulation and analytical cases. We note that the energy values are quite different but the maximum difference is the protein-protein interaction energy (Fig. \ref{passive_analytical}(g)), due to which, increasing $\rho$ is not very sensitive in our analytical model. Thus, we conclude that because of the simplicity of the analytical model, the quantitative comparison is not so good, however, it could describe the qualitative features of our system very well. Next, we compare the results for the active case.

In the active case, we compare the cross-sectional shape of the vesicle for a given $\rho$ and different values of $F$ in Fig. \ref{compare_active}. We note that the effect of increasing $F$ is very effective in analytical model. We also compare other results in Fig. \ref{active_analytical}. We note that the adhered fraction is very close to our MC simulations in the large $F$ region (Fig. \ref{active_analytical}(a)). In this case also, the energy values are not very comparable (see Fig. \ref{active_analytical}(c-g)).

\begin{figure}[ht]
\centering
\includegraphics[scale=0.75]{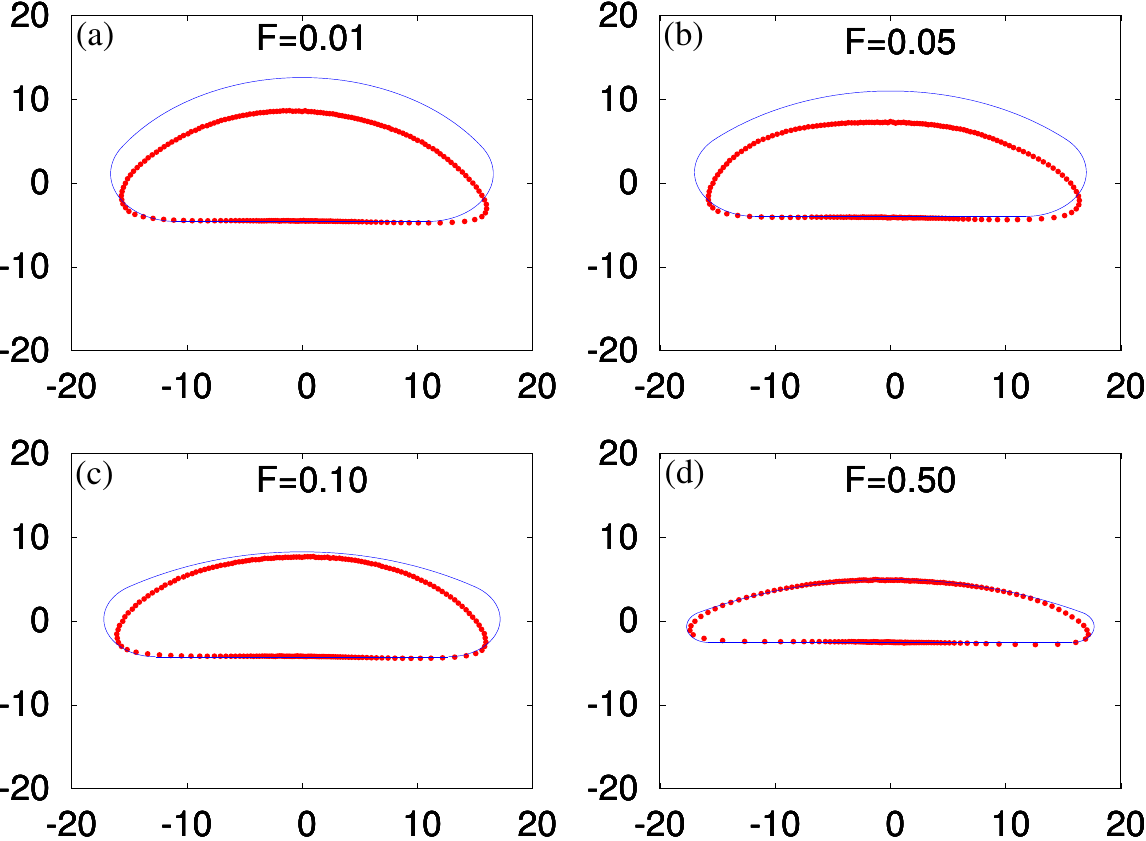}
\caption{Comparison of the shape of vesicle for simulation results and analytical results for active case with $\rho=10.36~ \%$, $E_{ad}=0.40 ~k_B T/l_{min}^2$ and various $F$. Red circles are for simulation results and blue line is for analytical results.}
\label{compare_active} 
\end{figure}

\begin{figure}[ht]
\centering
\includegraphics[scale=1.5]{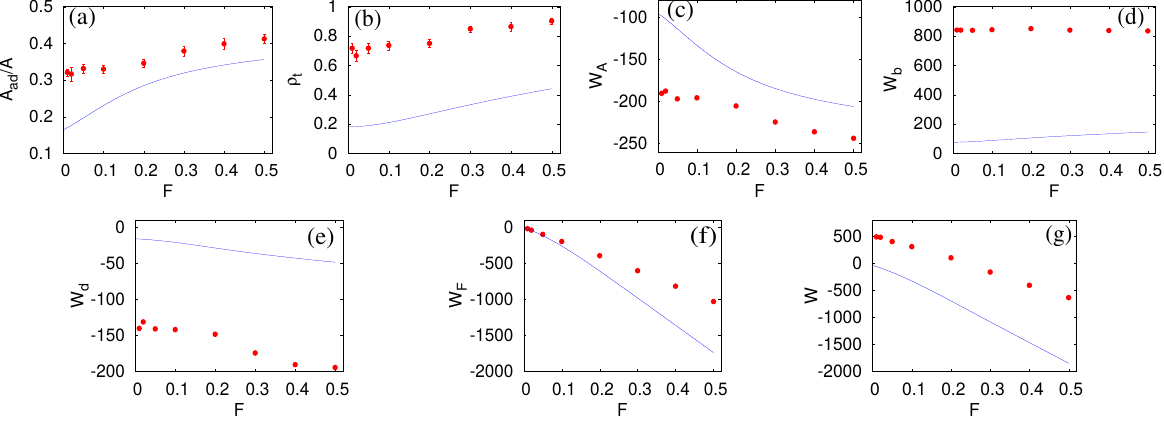}
\caption{Results for analytical model for active case and comparison with simulation. Red circles are for simulation results and blue line is for analytical results. (a)  Fraction of adhered area as a function of $F$. (b) The density of curved proteins in the annulus part of the vesicle ($\rho_t$) as a function of $F$. (c) The adhesion energy ($W_A$)) as a function of $F$. (d) The bending energy ($W_b$) as a function of $F$. (e) The protein-protein interaction energy ($W_d$) as a function of $F$. (f) The active energy due to cytoskeleton forces ($W_F$) as a function of $F$ (g) The total energy ($W$) as a function of $F$. Here we use $\rho=10.36 ~\%$ and $E_{ad}=0.40 ~k_B T/l_{min}^2$. Other parameters are same as Fig. \ref{passive_analytical}}
\label{active_analytical} 
\end{figure}

\section{Passive vesicle with proteins having zero spontaneous curvature}
\label{sec:passive_c0_0}
\begin{figure}[ht]
\centering
\includegraphics[scale=2.1]{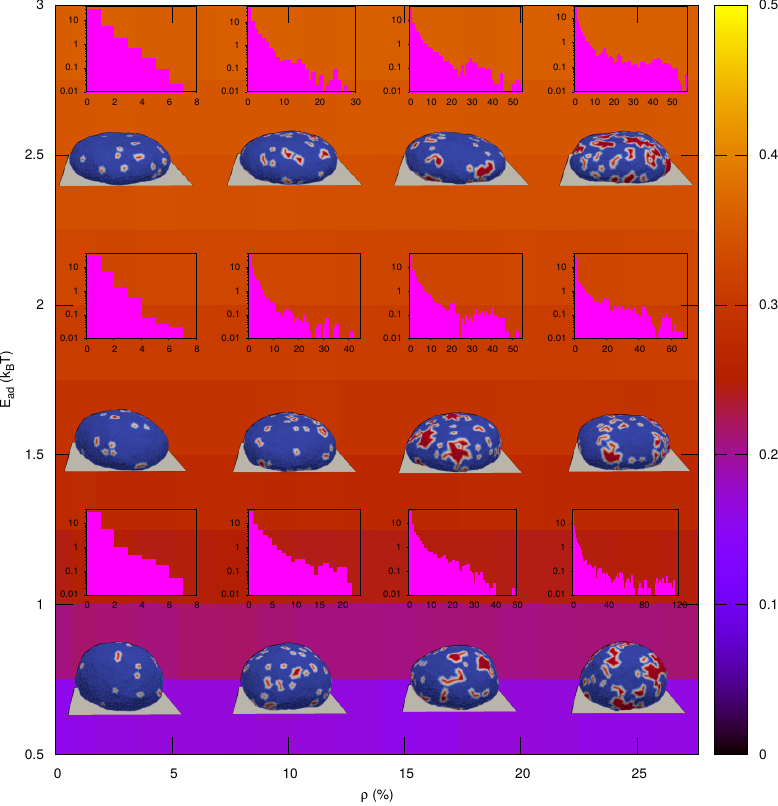}
\caption{Microstate of vesicle and cluster distributions for passive case with flat proteins ($c_0=0$). The background color is showing the fraction of adhered area. For the snapshots, we use $\rho=3.45 ~\%, 10.36~ \%,  17.27~\%, 24.18 ~\%$ and $E_{ad}= 0.75, 1.5, 2.5$ (in units of $k_B T$). The other parameters are the same as Fig. \ref{phase_passive}.}
\label{passive_zero_curvature} 
\end{figure}

In Fig.\ref{passive_zero_curvature} we show our results for a passive vesicle with proteins having zero spontaneous curvature, i.e., $c_0=0$. This serves as a verification of our calculation: since the proteins and the membrane are identical, we indeed find that the adhered area remains constant with the density of proteins, only depending on $E_{ad}$.  

\section{Linear stability analysis for budding transition and comparison with the contour for $<N_{cl}> = 2$}
\label{sec:Ncl2line}
In \cite{miha2019}, an expression for the critical temperature ($T^c$) is derived using linear stability analysis (Eq. 6 of \cite{miha2019}), below which the proteins will start forming aggregates (buds) \cite{Gladnikoff2009}. We use this expression and obtain the critical force $F^c$ as a function of $\rho$ and other variables as, $F^c = \frac{12 w}{l^2_{min} c_0} \frac{1}{(1-1/\rho R_0)} \Big(\frac{k_B T}{12 w \rho (1-\rho)} -1\Big)^2$. Here, $1/R_0$ is the mean curvature at the site of the curved membrane protein. In the limit $R_0 \rightarrow \infty$, the expression for $F^c$ will turn out to be,
\begin{equation}
F^c (R_0 \rightarrow \infty) \rightarrow \frac{12 w}{l^2_{min} c_0} \Big(\frac{k_B T}{12 w \rho (1-\rho)} -1\Big)^2
\label{eq:Fc}
\end{equation}

\begin{figure}[ht]
\centering
\includegraphics[scale=0.75]{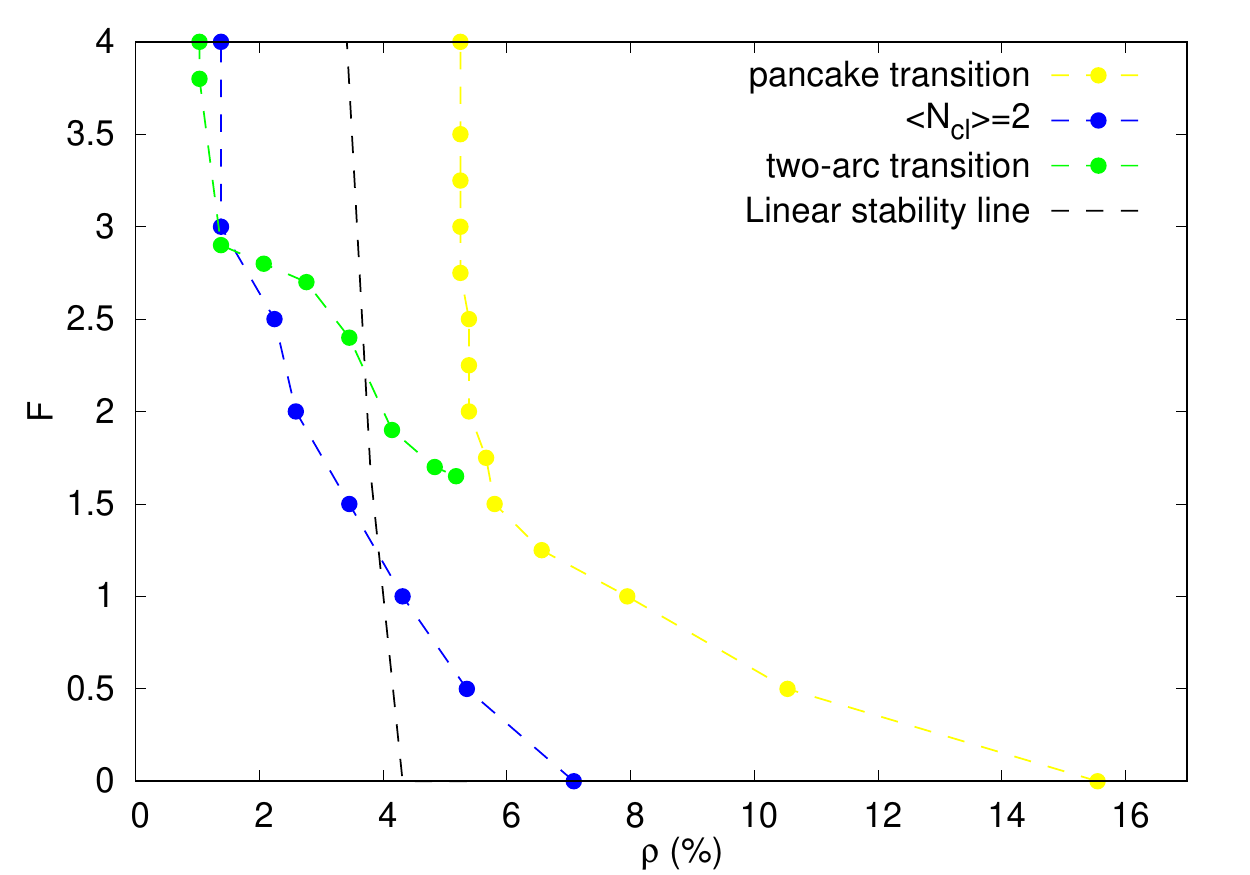}
\caption{The critical force  $F^c$ obtained by linear stability analysis (Eq.\ref{eq:Fc}) using the protein density at the contact line, and is compared with the line where $<N_{cl}>=2$. The other transition lines are as in Fig.\ref{phase_F_rho}.}
\label{Fc_linear_stability} 
\end{figure}

We use the above equation to calculate the $F^c$ for our case, and compare this line with the contour, where mean cluster $<N_{cl}> = 2$. Since, the budding forms along the contact line with the adhesive substrate, where the protein density is higher ($\rho_t$, say) than the average $\rho$ (see for example Fig.\ref{passive_cluster}a and appendix \ref{sec:analytical}), we use in Eq.\ref{eq:Fc} the value of $\rho_t$ obtained from the simulations to calculate the critical force $F^c$, and plot it as a function of the average $\rho$ (see Fig. \ref{Fc_linear_stability}). We note that the line is almost vertical, and in the large $F$ regime, it is in between the pancake transition (see Fig.\ref{phase_F_rho}) and the line $<N_{cl}> = 2$. Qualitatively, the shape of this analytic line, and its dependence on the density at the contact line, describe the transition of the proteins into small aggregates. The pancake transition at larger values of $\rho$ has qualitatively the same shape, but of course can not be captured by the linear stability analysis.

\section{Simulations with balanced total vertical force}
\label{sec:Fz_balance}
\begin{figure}[ht]
\centering
\includegraphics[scale=1.9]{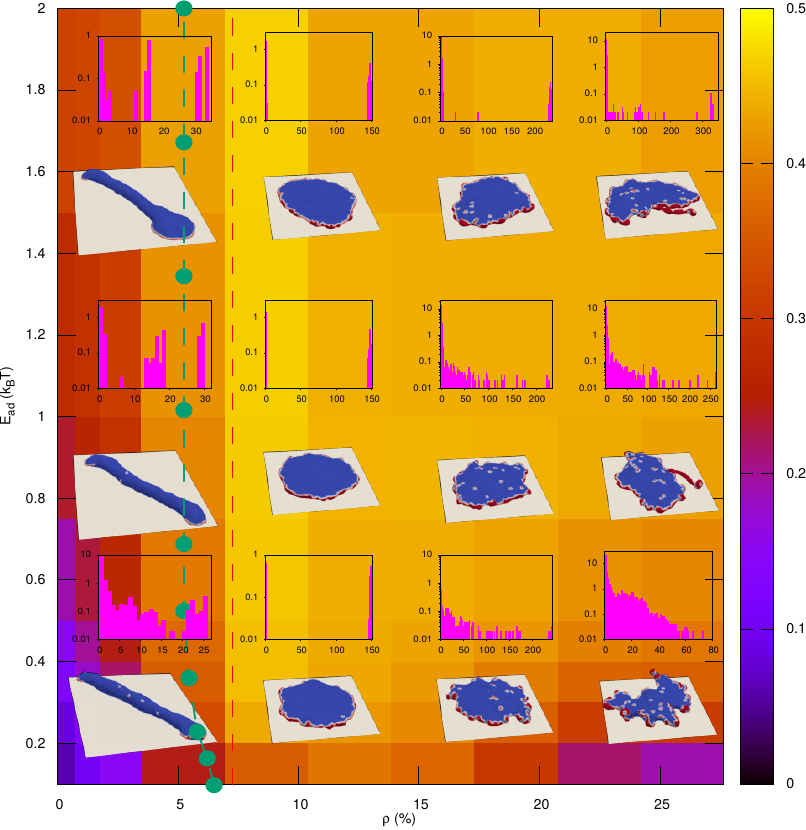}
\caption{Microstates and cluster distribution of vesicle for active case with balanced total vertical force. The external force is applied to all the vesicle nodes, such that it balances the net vertical force when is acts in the upward direction (\textit{i.e.,} when it acts against the adhesion process). We show the snapshots and cluster distribution of the vesicle with $E_{ad}=0.20, 0.75, 1.5$ (in units of $k_B T$) and $\rho$=3.45 \%, 10.36 \%, 17.27 \% and 24.18 \%. The green dashed line denotes the transition to a pancake-like shape. The red dashed vertical line denotes the density for the pancake transition for a free vesicle (without adhesion, \cite{miha2019}). Compared to the calculation without applying a balancing external force (Fig.\ref{phase_active}), we do not observe any qualitative change except for the fact that in this case, we could explore much smaller $E_{ad}$ values. Other parameters are the same as in Fig. \ref{phase_active}.}
\label{phase_active_FB} 
\end{figure}

The active force may in general act in a direction, that tries to de-adhere the vesicle from the adhesive surface. In order to cancel this effect, we apply an external force in the vertical direction (along $z$-direction), when the net vertical force acts upward (that tends to de-adhere the vesicle). We note that even after applying the external force, see Fig. \ref{phase_active_FB}, there is no qualitative change in the results. The benefit of applying this external force is that we could now explore much smaller $E_{ad}$ which could not be explored before, due to the de-adhesion of vesicle from the substrate.
\section{Active vesicle with zero spontaneous curvature proteins}
\label{sec:active_c0_0}
\begin{figure}[ht]
\centering
\includegraphics[scale=0.7]{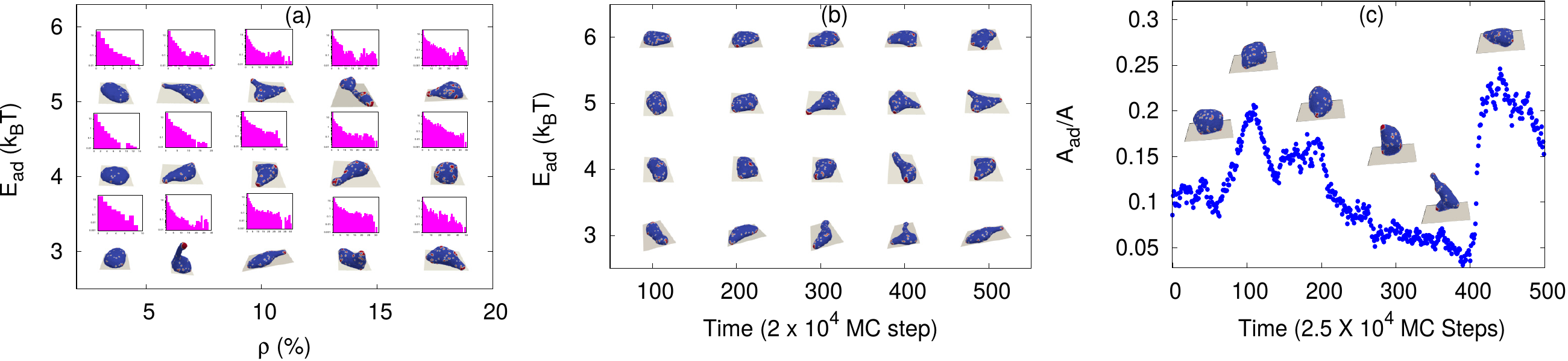}
\caption{Results for active case ($F=4~k_B T/l_{min}$) with proteins having zero spontaneous curvature. (a) Microstate of vesicle and cluster distributions for different values of $\rho$ and $E_{ad}$. We show the snapshots for $\rho=3.45 ~\%, 6.91 ~\% ,  10.36 ~\%, 13.82 ~\%, 17.27~\%$ and $E_{ad}= 3, 4, 5$ (in units of $k_B T$). (b) The dynamic nature of the vesicle for a given density $10.36 ~\%$ and  $E_{ad}=3, 4, 5, 6$ (in units of $k_B T$). (c) Variation of adhered fraction with time along with snapshots. We use here $\rho=10.36 ~\%$ and $E_{ad}=0.50 ~k_B T$. The other parameters are the same as in Fig.\ref{phase_active}.}
\label{active_c0_0} 
\end{figure}

In this section, we show our results for the active case with proteins having zero spontaneous curvature. The shape of the vesicle is very dynamic in this case and changes with time. We show few snapshots of the vesicle in Fig. \ref{active_c0_0}(a). The shapes shown in this figure should not be assumed to be a steady state shape. We also show the dynamic nature of the vesicle in Fig. \ref{active_c0_0}(b), where we show the snapshot for a given density ($\rho = 10.36 ~\%$) at different instant of time, for few values of $E_{ad}$. In Fig. \ref{active_c0_0} (c), we plot the fraction of adhered area with time, and also the snapshots near the plot. This plot also explain the dynamic nature of the vesicle.

\section{Spreading dynamics and time-dependent shapes of vesicles}
\label{sec:spreading_dynamics}
\begin{figure}[ht]
\centering
\includegraphics[scale=1.6]{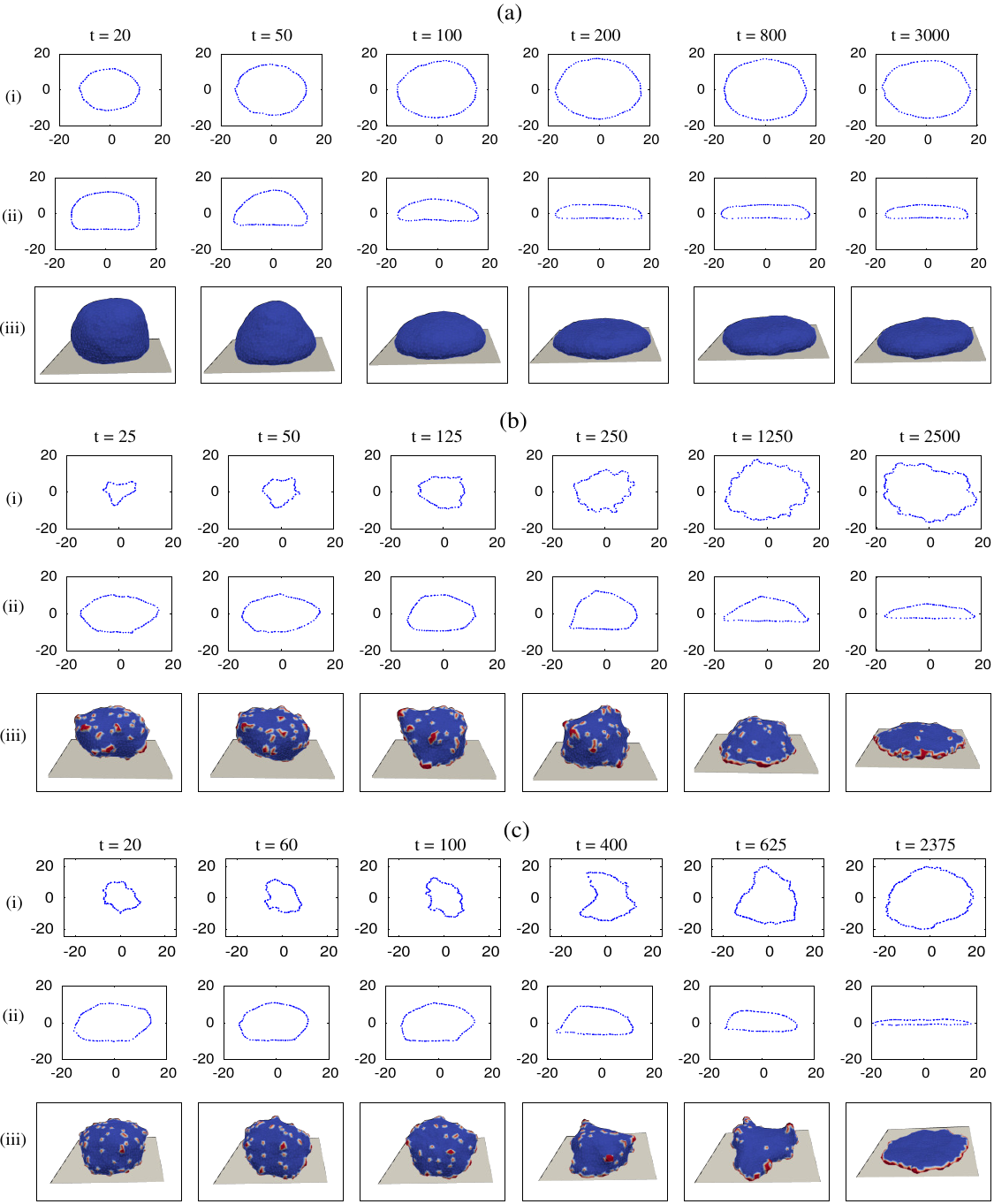}
\caption{Spreading dynamics. (a) Spreading of a protein free vesicle. Here, we choose $E_{ad}=5 ~k_B T$. (i) The bottom view of the adhered area with time, showing that the vesicle adhere quickly. (ii) The side view of the vesicle, showing how the cross-sectional area decreases with the adhesion of vesicles. (iii) Three-dimensional view of the vesicle with time. (b) Spreading of passive vesicle with $\rho=13.82 ~\%$ and $E_{ad}=0.50 ~k_B T$. Panels (i), (ii) and (iii) are the same as above. The adhered area grows uniformly from the beginning. (c) Spreading of an active vesicle, with $\rho=10.36 ~\%$, $E_{ad}=0.50 ~k_B T$ and $F=4 ~k_B T /l_{min}$. In the beginning, the growth of adhered area is slow, then accelerates and finally the vesicle takes the shape of a pancake. The unit of time (t) is taken as $2 \times 10^3$ MC steps.}
\label{spreading_dynamics} 
\end{figure}

Here, we show our results for the dynamics of spreading of the vesicles, by plotting the cross-sectional shapes and snapshots with time (Fig.\ref{spreading_dynamics}). We study three different cases: A protein free vesicle (Fig.\ref{spreading_dynamics} (a)), a vesicle with passive proteins  (Fig.\ref{spreading_dynamics} (b)) and a vesicle with active proteins  (Fig.\ref{spreading_dynamics} (c)). For the protein-free vesicle, we chose a large adhesion energy $E_{ad}=5.0 ~k_B T$ such that the vesicle could spread significantly. For the passive vesicle, we chose a small adhesion energy $E_{ad}=0.50 ~k_B T$ and $\rho=13.82~ \%$, in the regime where the curved proteins dominate the spreading dynamics. For the active case, we use $F=4 ~k_B T /l_{min}$, $E_{ad}=0.50 ~k_B T$ and a comparably smaller density of proteins $\rho=10.36 ~\%$, where the active forces dominate the spreading. For each of the three cases, we plot (i) the cross-section of adhered area, (ii) The side view of the vesicle, and (iii) The three-dimensional view of the vesicle. 

\section{Quantifying pancake transition by measuring the fraction of largest cluster}
\label{sec:largest_cluster}
\begin{figure}[ht]
\centering
\includegraphics[scale=1]{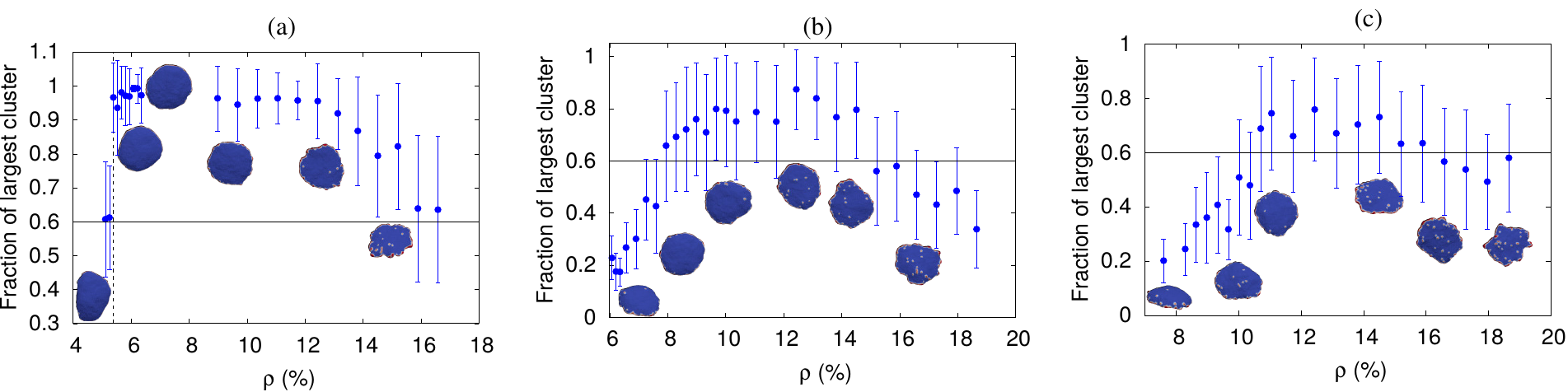}
\caption{Fraction of largest cluster with $\rho$ for various values of $F$ and $E_{ad}=0.50 ~k_B T$. (a) $F=2.0 ~k_B T/l_{min}$, (b) $F=1.0 ~k_B T/l_{min}$, and (c) $F=0.50 ~k_B T/l_{min}$. The vertical dashed line of (a) is showing the pancake transition, which is a sharp transition. For (b) or (c), the pancake transition is not very sharp, and the shape is assumed to be a pancake when the size of largest cluster is at least $60 ~ \%$ of the total number of proteins.}
\label{largest_cluster} 
\end{figure}

In Fig.\ref{largest_cluster} we show our results for the fraction of largest cluster as function of $\rho$ for active case with different values of $F$ and a small value of $E_{ad}=0.50 ~k_B T$. We note that for large $F$, there is a sharp jump in the value of the fraction at the pancake transition (dashed vertical line in Fig. \ref{largest_cluster} (a)), however, for small $F$, the fraction does not show any big jump. In order to quantify the pancake transition for small $F$, we choose a threshold value of the fraction, above which we assume the  shape to be a pancake. By monitoring the data of Fig. \ref{largest_cluster}, we choose the threshold value to be $0.60$, above which the shape looks like a pancake. 

\section{Adhered area-volume relation}
\label{sec:volume}
\begin{figure}[ht]
\centering
\includegraphics[scale=1.5]{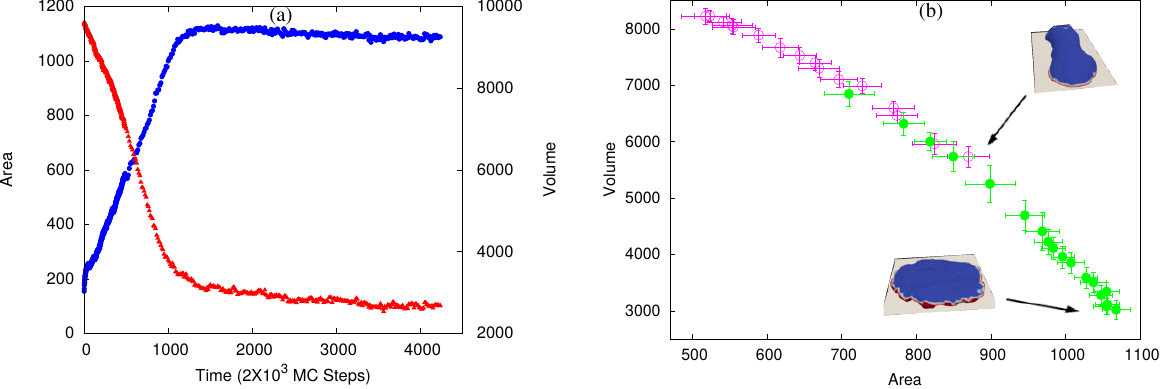}
\caption{Area-Volume relation. (a) The variation of area and volume with time for an active vesicle. The red triangles are for the volume and the blue circles are for the area. We use here, $\rho=10.36 ~\%$, $E_{ad}=0.50 ~k_B T$ and $F=4 ~k_B T /l_{min}$. (b) The area-volume relation in the steady state for active case with $E_{ad}=0.50~ k_B T$ and two different values of $\rho$. The hollow magenta circles are for $\rho=4.84 ~\%$ and the green solid circles are for $\rho=10.36 ~\%$. We also show the snapshot of vesicles for both the $\rho$, for their highest area. Here, we vary $F$ in order to access different area and volume in the steady state.  We note that for larger $\rho$ (which corresponds to the pancake shape), the system can access larger area and smaller volume in comparison to lower $\rho$ (which corresponds to the two-arc elongated shapes).}
\label{area_volume} 
\end{figure}

As the cells spread, their volume is observed to decrease \cite{guo2017cell,xie2018controlling}. We find similar dynamics in our model simulations. The steady-state relation between the vesicle volume and adhered area, is also similar to the experimental observations, although we have explored a smaller range of values (see Fig. \ref{area_volume}). Note that in cells there are internal organelles, such as the large nucleus, that limit the decrease in volume and which our model does not describe. In addition, osmotic pressure, and ion fluxes also contribute to the volume control in cells \cite{adar2020active}. 

\section{Speed of the crescent-shaped vesicle in the $E_{ad}-F$ plane }
\label{sec:crescent_speed}
\begin{figure}[ht]
\centering
\includegraphics[scale=0.7]{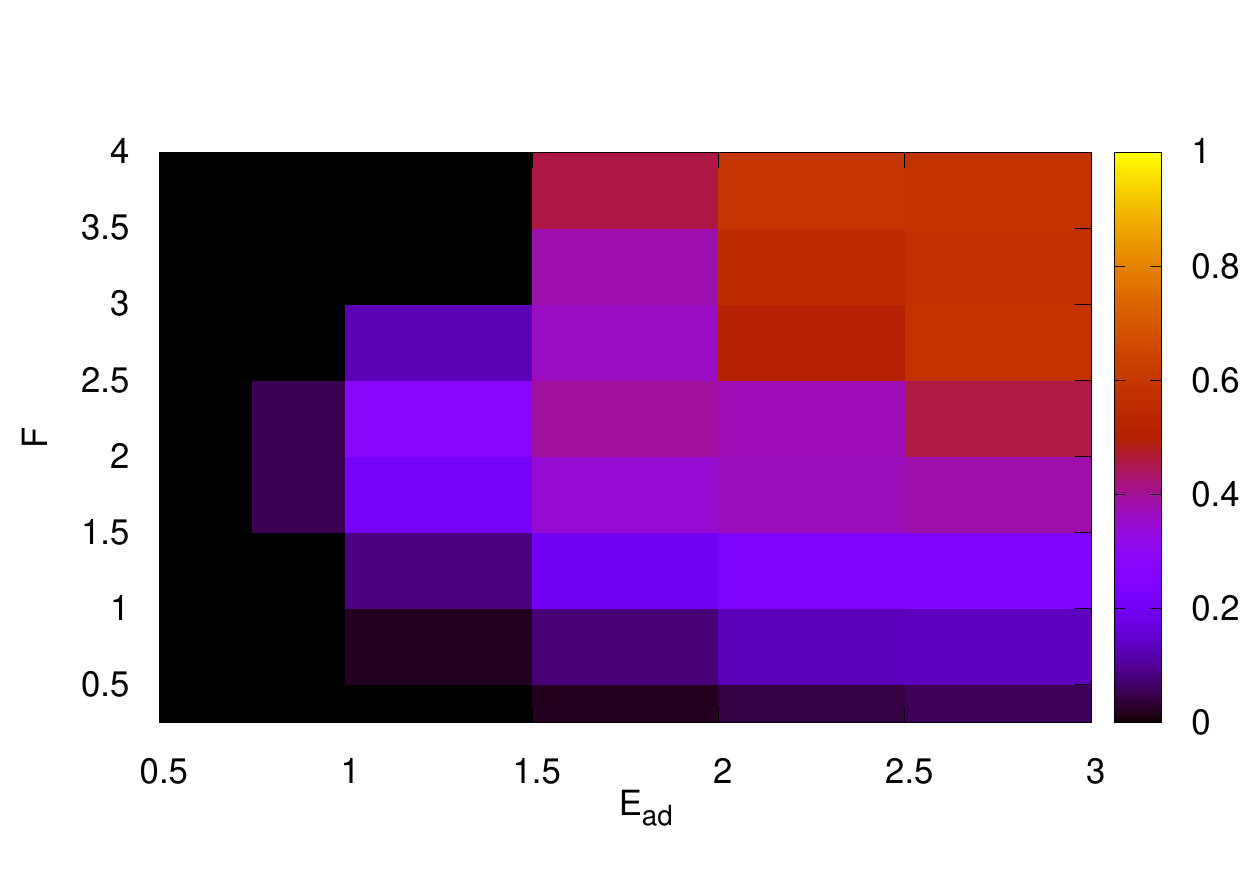}
\caption{Speed of the crescent-shaped vesicle in the $E_{ad}-F$ plane. Here, we calculate the speed as the displacement of COM per MC step and then normalize all the values by the maximum speed, where the maximum value of the speed is $0.328 ~ l_{min}$ per MC step. For large $E_{ad}$ but small $F$, the vesicle speed is zero but it might show diffusive behaviour as well. The speed is maximum for large $E_{ad}$ and large $F$ region.}
\label{crescent_speed} 
\end{figure}

In Fig.\ref{crescent_speed} we show the speed of the crescent-shaped vesicle in the $E_{ad}-F$ plane, without scaling by $E_{ad}$ (as is shown in Fig.\ref{phase_crescent}b). We normalize the speed by the maximum value, where the maximum value of the speed is $0.328 ~ l_{min}$ per MC steps. We note that, the speed is maximum for large $E_{ad}$ and large $F$ region.

\section{Mean square displacement of the crescent-shaped vesicle}
\label{sec:MSD}
\begin{figure}[ht]
\centering
\includegraphics[scale=0.9]{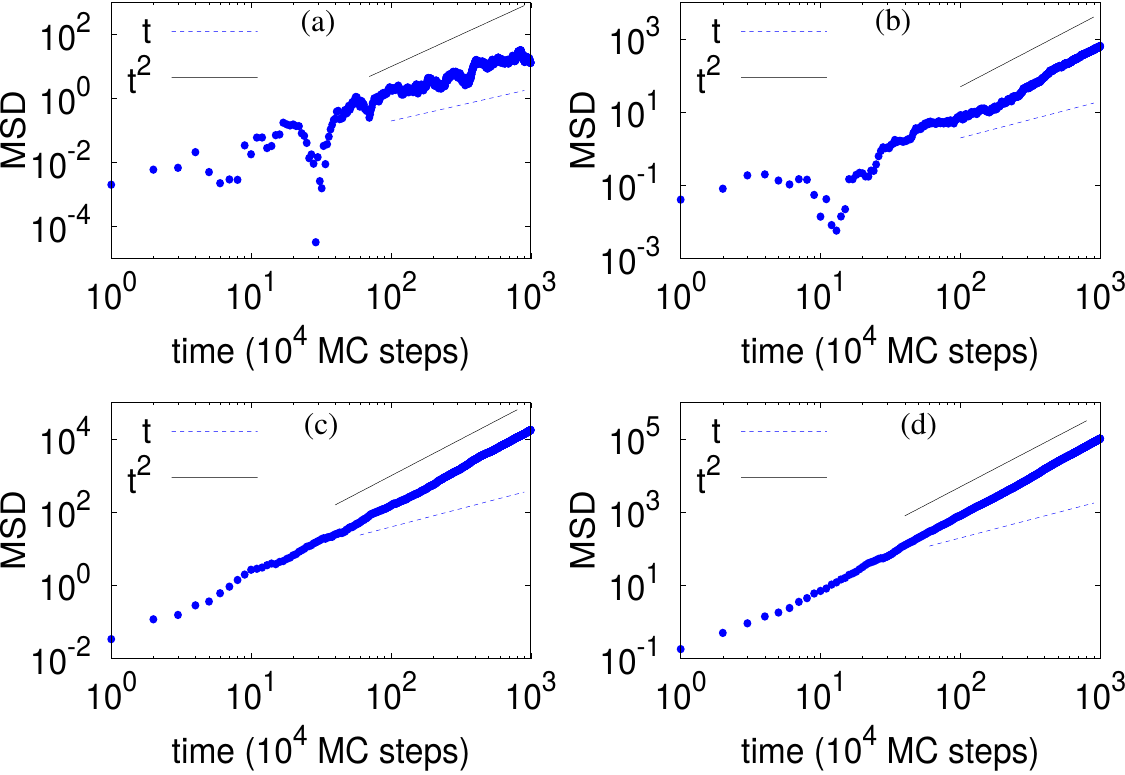}
\caption{Mean square displacement (MSD) of the COM of vesicle. (a) $F=0.25 ~k_B T/l_{min}$, (b) $F=0.50 ~k_B T/l_{min}$, (c) $F=2.0 ~k_B T/l_{min}$, and (d) $F=4.0 ~k_B T/l_{min}$. We note that for small $F$, MSD shows diffusive behaviour while for large $F$, the vesicle has a finite speed. The y-axis is in the unit of $l_{min}$, and the typical size of a vesicle (radius of a pancake shape, say) is $\sim 20 ~ l_{min}$. Here, we use $E_{ad}=2.0~ k_B T$, and $\rho=3.45 ~\%$.}
\label{MSD} 
\end{figure}

In Fig.\ref{MSD} we show the mean square displacement of the COM of the crescent-shaped vesicle with time, for $E_{ad}=2.0 ~k_B T$, $\rho=3.45 ~\%$ and various values of $F$. We note that for large $F$, there is a finite speed which increases with $F$, however, for small $F$, the displacement grows diffusively. 

\section{Effect of varying the protein-protein interaction strength $w$}
\label{sec:various_w}
\begin{figure}[ht]
\centering
\includegraphics[scale=1.5]{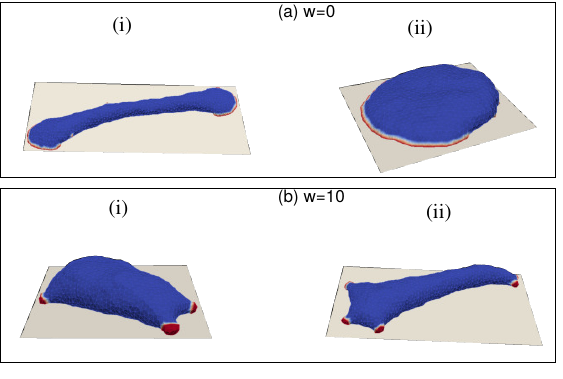}
\caption{Snapshots of vesicle with various values of protein-protein interaction strength $w$. (a) For $w=0 ~ k_B T$, we obtain the two-arc and crescent-like shapes as found for the case of $w=1~ k_B T$ (Fig.\ref{phase_crescent}). For (i), we use $E_{ad} = 2.0 ~ k_B T$, $F = 4.0 ~ k_B T/l_{min}$ and $\rho = 3.45 ~ \%$, and for (ii), we use $E_{ad} = 3.0 ~ k_B T$, $F = 2.0 ~ k_B T/l_{min}$ and $\rho = 3.45 ~ \%$. (b) For very large $w=10 ~ k_B T$, the vesicle forms small isolated clusters, that do not break-up or merge after long time. Here, for (i), we use $E_{ad} = 2.0 ~ k_B T$, $F = 3.0 ~ k_B T/l_{min}$ and $\rho = 3.45 ~ \%$, and for (ii), we use $E_{ad} = 2.0 ~ k_B T$, $F = 4.0 ~ k_B T/l_{min}$ and $\rho = 3.45 ~ \%$.}
\label{different_w} 
\end{figure}

Throughout the paper we kept the value of the protein-protein interaction strength ($w$) to be fixed at $w =1 ~k_B T$. In Fig.\ref{different_w}(a), we show that even if we take $w=0 ~ k_B T$, the qualitative nature of the shapes of vesicles do not change. We still obtain the two-arc shapes or the crescent-like shapes. However, increasing $w$ to a value such that $w\gg F\cdot l_{min}$ changes qualitatively the shapes of the vesicle, as shown for $w = 10 ~ k_B T$ in Fig.\ref{different_w})b. The bud-like protein clusters are now solid-like, and do not easily deform or break-up, and do not merge over time.

\section{Difference in the adhered area of a crescent-shaped vesicle and a two-arc shaped vesicle across the crescent to two-arc transition line (Fig.\ref{phase_crescent}(a))}
\label{sec:delta_A}
\begin{figure}[ht]
\centering
\includegraphics[scale=1.5]{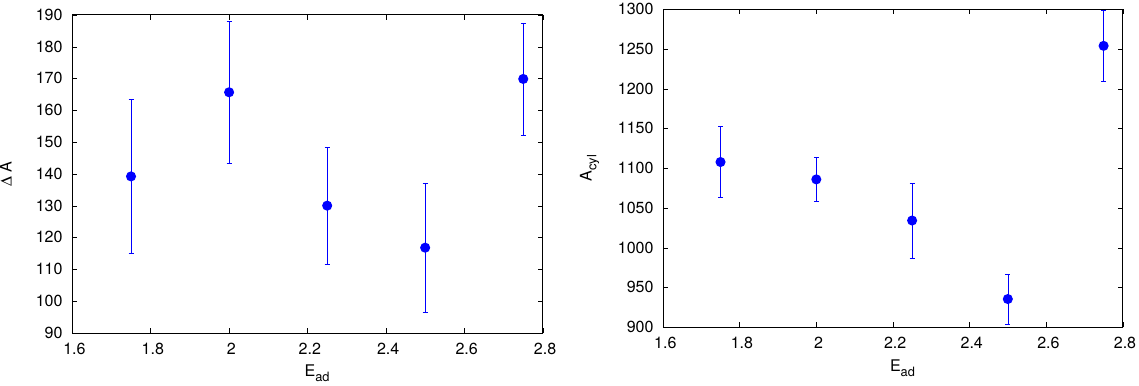}
\caption{(a) The difference in the adhered area of a crescent-shaped vesicle and a two-arc shaped vesicle across the crescent to two-arc transition line, $\Delta A$ (in units of $l^2_{min}$), as a function of $E_{ad}$. We note that the value of $\Delta A$ does not show any monotonic variation with $E_{ad}$, and is approximately constant along the transition line (as was assumed in Eq.\ref{crescentEad}). (b) The area of cylindrical part in a two-arc shape $A_{cyl}$ (in units of $l^2_{min}$), as a function of $E_{ad}$. This value also seems to be roughly a constant along the transition line.}
\label{delta_A} 
\end{figure}

The analytical calculation of the transition line (Fig. \ref{phase_crescent}(a)) separating a crescent shape and two-arc shape (Eq.\ref{crescentEad}) depends on the value of $\Delta A$, which is difference in adhered area of the crescent-shaped vesicle and the two-arc shaped vesicle across the transition line. Here, we plot the value of $\Delta A$ across this transition line as a function of $E_{ad}$ as extracted from the simulations (Fig. \ref{delta_A}(a)). The value of $\Delta A$ does not show any monotonic variation with $E_{ad}$, and is approximately constant along the transition line (as was assumed in Eq.\ref{crescentEad}). The average value of $\Delta A$ is $\simeq 144.32$. The analytical calculation also contains the area of the cylindrical part of the vesicle, $A_{cyl}$. We also measure this area, and note that this value is also roughly a constant along the transition line  (Fig. \ref{delta_A}(b)). The average value of this area turns out to be $A_{cyl} \simeq 1083.52$. We assume the number of proteins at each end of the cell to be $N=25$, half of the total number of proteins. Since, the unit of $E_{ad}$ is different in simulations and analytical calculation, we properly scale it to make it consistent for both the cases. Using all these values, we obtain the prefactor of $F^2$ in Eq.\ref{crescentEad} as $\simeq 1.96$ This value if very close to the value of $2.04$ obtained by fitting the simulation points along the transition line (Fig.\ref{phase_crescent}a) with the equation.

\section{Motile vesicle growing against an immobile barrier}
\label{sec:barrier}
\begin{figure}[ht]
\centering
\includegraphics[scale=0.75]{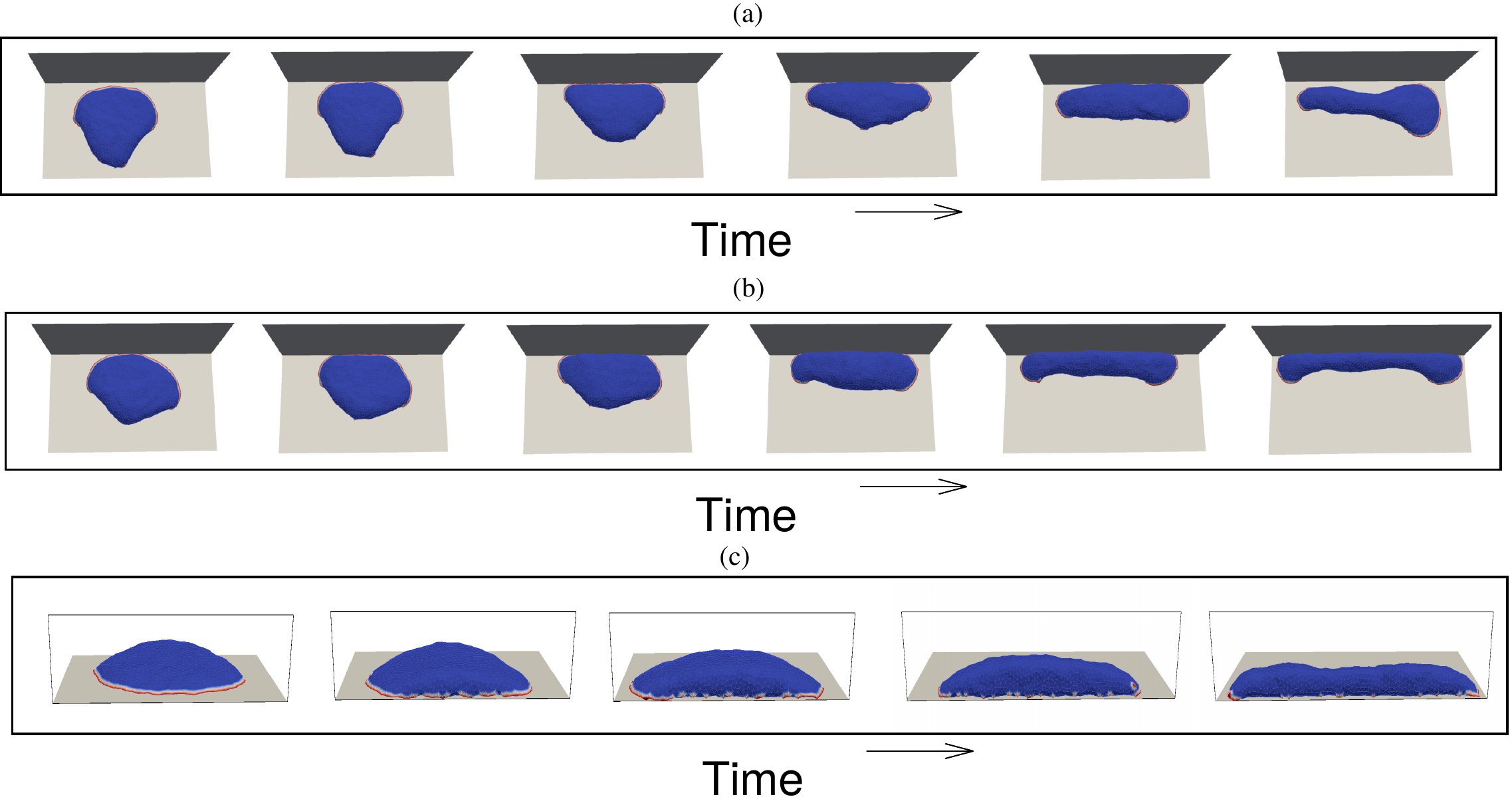}
\caption{Motile vesicle growing against an immobile barrier. (a) We plot the snapshots of vesicle with time for a non-adhesive barrier. (b) The snapshots of the vesicle with time for an adhesive barrier. (c) The side view of the vesicle for the adhesive case. We note that in both the adhesive and non-adhesive cases, the crescent-shaped vesicle breaks into two-arc shape. Here, we use $\rho=3.45 ~\%$, $F=4 ~k_B T/l_{min}$ and $E_{ad}=3.0 ~k_B T$. For the adhesive barrier also, we use $E_{ad}=3.0 ~k_B T$. }
\label{barrier} 
\end{figure}

In Fig.\ref{barrier} we allow a motile crescent shape to hit a rigid non-movable barrier, placed perpendicular to the direction of motion of the crescent. We take a non-adhesive as well as an adhesive barrier. In both the cases, the crescent shape finally breaks into two-arc shape. 


\end{document}